\documentclass[12 pt, journal, draftcls, onecolumn]{IEEEtran}

\usepackage{amsmath} 
\usepackage{amssymb}  
\usepackage{graphicx}
\usepackage{subfig}
\usepackage{tabularx}
\usepackage{multirow}
\usepackage{url}
\usepackage{setspace}
\usepackage{float}
\usepackage{cite}
\usepackage[table]{xcolor}
\usepackage{color}
\usepackage{soul}

\renewcommand{\baselinestretch}{1}

\title{\LARGE \bf
Sensor Fault Detection, Isolation and Identification Using Multiple Model-based Hybrid Kalman Filter for Gas Turbine Engines*}

\author{ Bahareh Pourbabaee$^{1}$, Nader Meskin$^{2}$ and Khashayar Khorasani$^{1}$ \\
\thanks{*This publication was made possible by NPRP grant No. 4 - 195 - 2 - 065 from the Qatar National Research Fund (a member of Qatar Foundation). The statements made herein are solely the responsibility of the authors.}
\thanks{$^{1}$ B. Pourbabaee and K. Khorasani are with the Department of Electrical and Computer Engineering, Concordia University, Montreal, Canada. {b\_pourba@ece.concordia.ca} and kash@ece.concordia.ca.}%
\thanks{$^{2}$ N. Meskin is with the Department of Electrical Engineering, Qatar University, Doha, Qatar. {nader.meskin@qu.edu.qa}.}
}
\begin{document}
\maketitle

\begin{abstract}
In this paper, a novel sensor fault detection, isolation and identification (FDII) strategy is proposed by using the multiple model (MM) approach. The scheme is based on multiple hybrid Kalman filters (HKF) which represents an integration of a nonlinear mathematical model of the system with a number of piecewise linear (PWL) models. The proposed fault detection and isolation (FDI) scheme is capable of detecting and isolating sensor faults during the entire operational regime of the system by interpolating the PWL models using a Bayesian approach. Moreover, the proposed multiple HKF-based FDI scheme is extended to identify the magnitude of a sensor fault by using a modified generalized likelihood ratio (GLR) method which relies on the healthy operational mode of the system.
To illustrate the capabilities of our proposed FDII methodology, extensive simulation studies are conducted for a nonlinear gas turbine engine. Various single and concurrent sensor fault scenarios are considered to demonstrate the effectiveness of our proposed on-line hierarchical multiple HKF-based FDII scheme under different flight modes. Finally, our proposed HKF-based FDI approach is compared with various filtering methods such as the linear, extended, unscented and cubature Kalman filters (LKF, EKF, UKF and CKF, respectively)
 corresponding to both \textit{interacting} and \textit{non-interacting} multiple model (MM) based schemes. Our comparative studies  confirm the superiority of our proposed HKF method in terms of  promptness of the fault detection, lower false alarm rates, as well as robustness with respect to the engine health parameters degradations.
\end{abstract}


\IEEEpeerreviewmaketitle
\vspace{-3mm}
\section{INTRODUCTION}
\label{Introduction}
An effective fault detection, isolation and identification (FDII) technology can play a crucial role in improving the system availability, safety and reliability as well as reducing the maintenance costs and risks of catastrophic failures. Over the past few years, many researchers have focused on proposing sophisticated fault detection and isolation (FDI) schemes constituting as a significant component of an FDII solution \cite{Isermann1997} and \cite{Frank1990}. The general methodology for an FDII consists of several important steps including: (a) generation of residuals as indicators of faults, (b) isolation of the faulty actuator, or sensor, or component element, and (c) identification and estimation of the severity of the fault parameter.

The FDII approaches can be categorized into three distinct groups including the model-based, data-based and hybrid methods \cite{Lin2011} and \cite{Zhang2013}. The model-based FDI approaches have been introduced in survey papers \cite{Frank1990, Isermann1997} and \cite{Venkatasubramanian2003, Hwang2010, Silva2012} and can be grouped into basic approaches, namely (i) Kalman filter and unknown input observer based approaches, (ii) parity relations, (iii) optimization-based algorithms, and (iv) parameter estimation and identification techniques.
Multiple model approach is a popular structure for FDII. It enables one not only to detect and isolate different faults but it also provides one with information on the magnitude of the fault and its identification information  \cite{Menke1995, Maybeck1999, Meskin2010, Meskin2013, Naderi2012, Pourbabaee2014}.

Since early research on FDII, gas turbines have been one of the challenging application areas which have received much attention. The main approach in gas turbines FDII is based on the Gas Path Analysis (GPA), which enables one to accomplish the actuator, sensor and component diagnosis by observing the engine's parameters such as the rotor speed, temperature and pressure at different stages and the fuel flow rates \cite{Volponi2003}. Various fault diagnosis techniques have been developed for gas turbines based on GPA ranging from Kalman filters \cite{Kobayashi2005, Kobayashi2003, Kobayashi2007, Simani2000}, neural networks \cite{Volponi2003theuse}, fuzzy logic \cite{Ganguli2003}, genetic algorithms \cite{Sampath2006}, sliding mode observer \cite{Rahme2015}, component adaptation approach \cite{Tsoutsanis2015} - \cite{Tsoutsanis2014}, and hybrid diagnosis \cite{Zhang2013}. Some of the linear approaches have been extended to fully nonlinear models of the engine that have led to nonlinear extensions of the Kalman filters to extended Kalman filters (EKF) and unscented Kalman filters (UKF) \cite{Naderi2012} and \cite{Shang2011}.

In this paper, a modular and a hierarchical non-interacting MM-based approach is proposed for both single and concurrent permanent sensors fault detection and isolation in a single spool jet engine. In our proposed hierarchical approach, multiple levels of detection filters are used to detect and isolate concurrent faults where at each time only one level of the scheme is active according to the engine sensors health status. 

We also propose a \emph{hybrid Kalman filter} (HKF) \cite{Kobayashi2007} scheme that consists of a single nonlinear on-board engine model (OBEM) augmented with piecewise linear (PWL) models constituting as the MM-based estimators to cover the \textit{entire engine operating} regime. Also, the discrete-time derivation of the HKF is formally shown in this paper. Therefore, multiple HKFs are constructed to ensure that the FDII algorithm works effectively in a wide range of operating conditions by decomposing the engine operating range into sub-regions each represented by a PWL model. We then apply a Bayesian approach to generate a general combined model based on the PWL models normalized weights. This provides us with a soft transition or interpolation among the PWL models. At any given operating point, the combined model represents the nonlinear system dynamics with less error than a single PWL model.

For interpolating the PWL models, various techniques are available in the literature including Takagi-Sugeno fuzzy models \cite{Babuska1997}, the functional state approach using discrete representations such as automaton for describing the transitions between regions \cite{Zhang1994}, Markov mixtures of experts to define the transitions \cite{Meila1997}, etc. Unlike the functional state approach, the Bayesian and fuzzy approaches enable soft transitions among the PWL models as the operating condition changes between various points. Moreover, the Bayesian approach uses the statistical method to infer the applicable and most appropriate operating regime at each time instant.

Sensor fault severity identification is another important part of our proposed FDII framework which may either be utilized for reconstructing the correct sensor measurements to design an output feedback control strategy or to remove the faulty sensor from the measurement set. The reconstructed fault-free system outputs are essential for a normal closed-loop system operation with a faulty sensor in order to maintain the system stability and set-point tracking under faulty situations \cite{Seron2013}. Towards this end, we have integrated the generalized likelihood ratio (GLR)-based method \cite{Willsky1976, Prakash2005, Patwardhan2006} with our MM-based scheme to estimate the sensor fault severity under various single and concurrent fault scenarios. The MM-based scheme does provide us with the time and location of a sensor fault that subsequently simplifies the use of the GLR method without requiring to solve an optimization problem for determining the fault detection time as part of its fault estimation process.

One of the major non-fault related factors leading to false alarms in a given FDI scheme arises due to the normal aging effects of the engine components such as compressor fouling, turbine erosion and wear or buckling of a component. The above phenomena are associated with gradual degradation of the health parameters from their healthy reference baselines, which is a type of modeling uncertainty. Different approaches, including the robust generation and evaluation methods, are introduced in \cite{Frank1997} to improve the robustness of the on-line FDI algorithm with respect to modeling uncertainties. In \cite{Seliger1991}, modeling uncertainty is assumed as the structured input with known distribution matrix and certain rank conditions in order to design a decoupling filter to achieve the robustness. However, in most applications including ours, usually it is challenging to model the uncertainty in a structured format. Therefore, robust residual evaluation methods in which the adaptive threshold is designed based on the upper bound of the unstructured modeling uncertainty that use the historical data have been introduced. Nevertheless, it is required to design an individual threshold for each fault type. In addition, in few works such as \cite{Zhang2013}, a periodic updating mechanism is used for the on-board engine model in order to enhance the robustness and the fault sensitivity.

In our paper, it is assumed that the reference baselines for the engine health parameters, namely the efficiencies and the mass flow rates of the turbine and the compressor, can be estimated after a certain number of flights by using an off-line health monitoring module \cite{Simon2008} for determining the percentages of degradations. The applied method can be utilized off-line in a ground station given that the health parameters are degraded more slowly as compared to an abrupt fault. Consequently, in order to ensure the reliability of our proposed FDII scheme during the engine life cycle, the reference baselines are periodically updated for the OBEM health parameters. This procedure will be useful in preventing occurrence of false alarms due to the engine health parameters degradations. Despite this updating process, there is always a mismatch between the values of the real engine health parameters and the ones that are used in the OBEM due to the off-line estimation errors of the health monitoring method. Therefore, we will also investigate the robustness of our proposed FDII scheme with respect to various estimation errors through performing Monte Carlo simulations.

To summarize the \underline{main contributions} of this paper can be stated as follows:
\begin{enumerate}
\item The explicit derivations of the discrete-time HKF scheme is formally provided in Appendix \ref{Appendix A} to complement the continuous-time HKF scheme derivations provided in \cite{Kobayashi2007}.
\item A modular and hierarchical non-interacting MM-based HKF structure is developed to detect and isolate single and concurrent sensor faults during the entire engine operating regime (flight profile) having lower fault detection time and better robustness towards the engine health parameters degradations as compared with the other linear and nonlinear filtering methods in the literature. Our method is capable of operating during the entire engine life cycle through periodically updating the reference baselines of the engine health parameters for the nonlinear OBEM. In addition, the computational time of our proposed method is lower than the investigated linear and nonlinear methods.
\item Through performing extensive simulation studies and measuring the mean of residual signals, it is shown that the HKF is capable of estimating the engine outputs more accurately with less number of operating points and false alarm rates as compared to multiple linear Kalman filters (MLKF) based FDI method.
\item Our proposed MM-based HKF structure is integrated with the GLR scheme to estimate the fault severity. This is accomplished by specifically eliminating  the estimation of the fault detection time which is a necessary step in the standard GLR scheme.
\item Our proposed multiple HKF-based FDI scheme is compared with the MM-based schemes that utilize various linear and nonlinear filtering approaches such as the linear Kalman filter (LKF), extended Kalman filter (EKF), unscented Kalman filter (UKF) and the cubature Kalman filter (CKF) in terms of the promptness of the fault detection, false alarm and incorrect fault detection rates, robustness with respect to the engine degradations and computational time.
\end{enumerate}

This paper is the extended version of \cite{Pourbabaee2013} from both theoretical and case-based simulation scenarios perspectives. Theoretically, we formally show how the HKF is derived and how the previously proposed sensor FDI algorithm in \cite{Pourbabaee2013} is integrated with the GLR method to investigate the sensor fault identification problem. The case study scenarios are also extended from the following perspectives, namely (a) different single and concurrent sensor faults are considered, (b) the reductions in the fault detection times, lower false alarms and incorrect fault detection rates under various parameter uncertainty levels are investigated, and (c) our proposed FDI scheme is compared with  linear and nonlinear observer-based methods in the literature (LKF, EKF, UKF, and CKF) in terms of  fault detection time, robustness towards the engine health degradations and the computational time.

The remainder of the paper is organized as follows. Section II provides the design of the HKF which includes the OBEM and the PWL models. The interpolation of the PWL models is provided in Section III, and Section IV describes the overall hierarchical MM-based FDI scheme applicable to both single and concurrent sensor faults. The modified GLR-based approach is developed and described in Section V for estimating the severity of a sensor bias fault. A number of Monte Carlo simulation scenarios and cases associated with various faults under multiple operational conditions are presented. Finally,  comparisons with various linear and nonlinear filtering methods (LKF, EKF, UKF and CKF) are included in Section VI, and  the paper is concluded in Section VII.

\section{HYBRID KALMAN FILTER (HKF) DESIGN}
\label{HKF}
An actual aircraft gas turbine engine,  used for deriving the on-board engine model (OBEM), and which is used for on-line diagnostic analysis can be described according to the following representation:
\begin{align}
& \dot{X}(t)= \mathcal{F}_c(X(t),H(t),U(t),\zeta(t)), \notag \\
& Y(t)= \mathcal{G}_c(X(t),H(t))+v(t),
\label{1}	
\end{align}

\noindent where $X(t) \in \mathbb{R}^n$, $H(t) \in \mathbb{R}^r$, $Y(t) \in \mathbb{R}^q $, $U(t) \in \mathbb{R}^p$, $\zeta(t)\in \mathbb{R}^2$ and $v(t) \in \mathbb{R}^q$ denote the engine state variables, health parameters, sensor measurements, input signals, and Gaussian zero-mean process and measurement noise at time $t$, respectively. The engine dynamics is also an implicit function of the ambient condition parameters including the ambient temperature and pressure. The ambient parameters are defined in terms of the environmental parameters including the altitude and the Mach number. The engine health parameters in \eqref{1} will become degraded from their healthy reference baselines during the entire engine life cycle. Moreover, it is assumed that the inputs and outputs of an actual engine are discretized with sufficiently small sampling period for performing simulation and implementation of our proposed on-line fault detection, isolation and identification (FDII) scheme. The discrete-time representation of the variables in \eqref{1} are denoted by $X(k)$, $Y(k)$, $H(k)$, $U(k)$, $\zeta(k)$ and $v(k)$ with the discrete dynamic functions $\mathcal{F}$ and $\mathcal{G}$ replacing $\mathcal{F}_c$ and $\mathcal{G}_c$, respectively.

The hybrid Kalman filter (HKF) consists of two main blocks that include a nonlinear on-board engine model (OBEM) and multiple piecewise linear (PWL) models derived at different operating points to cover the entire engine operating range. The continuous-time fault-free representation of the OBEM which can also be derived based on thermodynamics laws is now given as follows:
\begin{align}
&\dot{X}_{\text{OBEM}}(t)=f_c(X_{\text{OBEM}}(t),H_{\rm{OBEM}}(t),U(t)), \notag \\
&Y_{\text{OBEM}}(t)=g_c(X_{\text{OBEM}}(t),H_{\rm{OBEM}}(t)),
\label{2}
\end{align}

\noindent where $X_{\rm OBEM}(t)\in \mathbb{R}^n $ and $Y_{\rm OBEM}(t)\in \mathbb{R}^q $ denote the OBEM state variables and outputs. Both the actual engine and the OBEM operate in parallel under the same flight conditions. Moreover, $H_{\rm OBEM}(t) \in \mathbb{R}^r$ denotes the OBEM health parameters that can be represented by $H_{\text{OBEM}}(t)=\lambda^{\rm T} h(X_{\text{OBEM}}(t))$, in which $\lambda$ defines the OBEM health parameters degradation factors (reference baselines). This parameter  is an all-ones vector for the OBEM that represents a healthy or non-degraded engine. It is also assumed that the OBEM health parameters can be periodically updated and $\lambda$ is considered as fixed in between the updating intervals. In addition, $h(X_{\text{OBEM}}(t))$ denotes a state-dependent smooth function that corresponds to the compressor and the turbine performance maps and is modeled as a polynomial function for our developed gas turbine engine model. The updating process for the OBEM health parameters will be described below in detail. Moreover, for performing simulation and implementations, the continuous-time OBEM model is assumed to be discretized with sufficiently small sampling period. The discrete-time representation of the variables in \eqref{2} are denoted by $X_{\rm OBEM}(k)$, $Y_{\rm OBEM}(k)$, $H_{\rm OBEM}(k)$ and $U(k)$ with the discrete dynamic functions $f$ and $g$ replacing $f_c$ and $g_c$, respectively.

The continuous-time OBEM model is linearized and discretized at multiple operating points (corresponding to engine steady-state values) that are denoted by $(X_{{\rm{ss}}_i},U_{{\rm{ss}}_i},Y_{{\rm{ss}}_i})$ with sufficiently small sampling period. The constructed multiple linear discrete-time state-space models is now given by:
\begin{align}
& \Delta X_i(k+1)= A|_{X_{{\rm{ss}}_i}}\Delta X_i(k)+B|_{U_{{\rm{ss}}_i}}\Delta U_i(k), \notag \\
& \Delta Y_i(k)= C|_{X_{{\rm{ss}}_i}}\Delta X_i(k),
\label{3}
\end{align}

\noindent where $i \in {1,\ldots, L}$ ($L$ is the number of the operating points), $A|_{X_{{\rm{ss}}_i}}$, $B|_{U_{{\rm{ss}}_i}}$ and $C|_{X_{{\rm{ss}}_i}}$ denote the state-space matrices associated with the $i^{th}$ operating point, and $\Delta X_i(k)=X_{\rm{OBEM}}(k)-X_{{\rm{ss}}_i}$, $\Delta Y_i(k)=Y_{\rm{OBEM}}(k)-Y_{{\rm{ss}}_i}$ and $\Delta U_i(k)=U(k)-U_{{\rm{ss}}_i}$. In the Appendix A, the linearization and discretization process of the continuous-time OBEM model is described in more detail. In this paper, it is assumed that the OBEM linearized models are obtained with the health parameters that are set to their healthy reference baselines. Also, the health parameter effects due to changes in the engine state variables have been implicitly incorporated into the matrices $A$ and $C$, although the deviations from the healthy reference baselines are not incorporated in the linear models.

For each linearized model, an off-line linear Kalman filter is designed to estimate both the actual engine states and sensor outputs as follows:
\begin{align}
&\Delta\hat{X}_i(k+1)=A|_{X_{{\rm{ss}}_i}}\Delta\hat{X}_i(k)+B|_{U_{{\rm{ss}}_i}}\Delta U_i(k)+K_{\rm ss}^i(Y(k)-\hat{Y}_i(k)), \notag\\
&\Delta\hat{Y}_i(k)=C|_{X_{{\rm{ss}}_i}}\Delta\hat{X}_i(k),
\label{4}	
\end{align}

\noindent where $\Delta \hat{X}_i(k)=\hat{X}(k)-X_{{\rm{ss}}_i}$, $\Delta \hat{Y}_i(k)=\hat{Y}(k)-Y_{{\rm{ss}}_i}$ and $K_{\rm ss}^i$ denotes the steady-state Kalman filter gain matrix. For the purpose of constructing the multiple model HKFs, the steady-state Kalman gain matrices as well as the matrices $A|_{X_{{\rm{ss}}_i}}$ and $C|_{X_{{\rm{ss}}_i}}$ that are constructed associated with multiple operating points are stored in a look-up table. According to \eqref{4}, the linear Kalman filter does not take into account the effects of the health parameter degradations from their healthy reference baselines since the state-space matrices have already been determined for an all-ones $\lambda$. Consequently, it does not have the required level of robustness for handling a vast number of health degradations that occur during the entire engine life cycle. Therefore, it is essential to update the health parameter reference baselines of the OBEM to maintain the reliability and accuracy of the state estimates and the performance of the FDII scheme through out the \textit{entire engine life cycle} operation. For this purpose, the health parameter reference baselines can be estimated by an off-line health monitoring system and then periodically updated in the discrete-time OBEM model. Therefore, the on-line FDII scheme is integrated with an off-line health monitoring system.

The frequency of the health monitoring system updates is significantly lower than that of the on-line FDII algorithm, since the health degradation process is  slow and gradual during one flight, although their accumulated effects after a number of flights may generate a large discrepancy between the OBEM and the actual engine. The off-line health monitoring module can either be a single augmented Kalman filter that estimates the health parameters based on the collected data during several flights or a nonlinear approximation method such as a neural network that receives the engine historical data. It must be noted that both the health monitoring and updating mechanism can be performed on-line having a sufficiently large sampling interval. However, the health parameter estimation process needs to be terminated whenever a fault occurs in the engine to avoid generating incorrect estimates of the health parameters.

The off-line health monitoring module uses the collected input and the measured output data to estimate the health parameter reference baselines, that is $\hat{\lambda}$. The estimated $\hat{\lambda}$ is updated periodically in the OBEM model within a determined time interval $\mathcal{T}$ that is usually a given number of flights or days. The update process feeds $\hat\lambda$ into the OBEM model so that the on-line FDII scheme can operate within the neighborhood of the degraded actual engine condition. Consequently, the OBEM model can now be re-written as follows:
\begin{align}
&X_{\text{OBEM}}(k+1)=f(X_{\text{OBEM}}(k),\hat{\lambda} h(X_{\text{OBEM}}(k)),U(k)), \notag \\
&Y_{\text{OBEM}}(k)=g(X_{\text{OBEM}}(k),\hat{\lambda} h(X_{\text{OBEM}}(k))).
\label{5}	
\end{align}

It must be noted that the OBEM model that is provided in \eqref{2} is used only once for performing the linearization process with $\lambda$ set to an all-ones vector, although the OBEM model in  \eqref{5}, with its reference baselines  periodically updated, is the one that is utilized in our proposed HKF structure as well as in our on-line FDII scheme. For sake of notational simplicity, we use the same notations for the above two versions of the OBEM model.

To construct the HKF representation we modify \eqref{4} where the steady-state variables are replaced by the OBEM states and outputs that are obtained from \eqref{5} and also by using the previously stored steady-state Kalman filter gain and state-space matrices as follows:
\begin{align}
&\hat{X}_i(k+1)-X_{\rm{OBEM}}(k+1)=A|_{X_{{\rm{ss}}_i}}(\hat{X}_i(k)-X_{\text{OBEM}}(k))+K_{\rm ss}^i(Y(k)-\hat{Y}_i(k)), \notag\\
&\hat{Y}_i(k)=C|_{X_{{\rm{ss}}_i}}(\hat{X}_i(k)-X_{\text{OBEM}}(k))+Y_{\text{OBEM}}(k).
\label{6}
\end{align}

\noindent In the above model the effects of the input and the $B$ matrix are eliminated from the HKF formulation given the fact that these have already been accounted for by the OBEM model. The procedure for derivation of the discrete-time HKF is formally shown in the Appendix A. 

Consequently, multiple HKFs are designed by using \eqref{6} for multiple operating points.
After updating the OBEM health parameter reference baselines, there is no longer a need to recalculate the $A|_{X_{{\rm{ss}}_i}}$, $C|_{X_{{\rm{ss}}_i}}$ and $K_{\rm ss}^i$ matrices for each operating point, since the effects of the health parameter degradations have been incorporated into the $X_{\rm OBEM}$ and $Y_{\rm OBEM}$ as given by \eqref{5}.

One of the factors that can affect the efficiency of our proposed FDII algorithm is the estimation error of the off-line health monitoring module, that leads to mismatches between the actual engine and the OBEM outputs.
Larger estimation errors may increase the fault detection time, and lead to occurrence of false alarms and incorrect fault detection rates. The acceptable ranges of the estimation errors that do not lead to false alarms and incorrect fault detection rates will be specified subsequently in Section \ref{SIMULATION RESULTS} for different health parameters under various healthy and faulty scenarios. Moreover, the reliability of our proposed on-line FDII strategy as a function of different mismatching factors between the actual operational engine and the OBEM (as represented by the reference baseline estimation errors $({\rm RBEE}=\frac{|\lambda-\hat\lambda|}{\lambda})$) and the magnitude of the process and measurement noise signals will be investigated subsequently in Section \ref{SIMULATION RESULTS} by means of a confusion matrix analysis.

In order to develop our proposed HKF-based scheme as an FDII strategy, multiple piecewise linear (PWL) models need to be generated for each fault hypothesis at various operating points. Moreover, the PWL models will be integrated and fused to cover the entire operational regime of an engine. The detail description of this process is provided below.

\section{Piecewise Linear Models (PWL) Interpolation}
\label{PWL}
Any given linear model of a nonlinear system has a limited operating range in which it remains valid. Nevertheless, our ultimate goal is to obtain a globally valid model which is valid for the entire operating regime. Therefore, the full operating range is divided into several sub-regions where each is defined around an operating point for which a piecewise linear (PWL) model can be derived \cite{Murray1997}. The PWL models can then be integrated in order to construct a parameter-varying general model whose parameters are the PWL models weights that are obtained through an on-line Bayesian approach. This will provide one with a soft interpolation among the PWL models as opposed to a hard switching among them. In this paper, the engine inputs including the fuel flow rate and ambient variables are used to partition the engine operational regime into multiple operating points for which the PWL models are constructed. These operating points are associated with different flight conditions such as take-off, climbing, cruise and landing modes. The selected number of operating points depends on (a) the required HKF state estimation accuracy, (b) the FDII strategy reliability on correct decisions, and (c) false alarm rates within the range of the applied health parameter degradations. Hence, if there are no concerns on the memory utilization, the number of the operating points can be selected to be as high as possible to enhance the HKF estimation accuracy and the FDII scheme valid decision rates and also to decrease the false alarm rates.

One of the important advantages of our proposed HKF scheme is in requiring a smaller number of operating points as compared to standard linear Kalman filters for covering an entire operational regime of the engine. This is facilitated and made possible due to substitutions of the steady-state variables in \eqref{6} by the OBEM state and output variables. The operating range of a PWL model in  \eqref{4} is only limited to the neighborhood of a corresponding operating point, although this can be extended to a larger range in  \eqref{6} given that $X_{\rm OBEM}$ and $Y_{\rm OBEM}$ are changed according to the engine operating condition, which enable the PWL model to be valid in a wider range.

Since, the sensor fault is injected into the actual gas turbine engine and not the OBEM, we have for the faulty engine
\begin{align}
&X(k+1)= \mathcal{F}(X(k),H(k),U(k),\zeta(k)), \notag\\
&Y(k) = \mathcal{G}(X(k),H(k))+\sum_{s=1}^{q}b_s z_s\delta(k-k_{fs})+v(k),
\label{7}	
\end{align}
\noindent where $\mathcal{F}$ and $\mathcal{G}$ represent the discrete-time dynamic equations of the actual gas turbine engine, $q$ is the number of sensors, $b_s$ represents the $s^{th}$ sensor bias fault magnitude and $z_s$ represents the fault location vector that has a unit value for the $s^{th}$ element while the other elements are set to zero, and $\delta(k-k_{fs})$ denotes a unit step function that occurs at the sample $k_{fs}$ corresponding to the $s^{th}$ fault occurrence time. The bias is set to zero for the healthy sensor scenario. Therefore, there are a total of  $q+1$ sensor modes (corresponding to one healthy and $q$ faulty sensor modes).

The PWL models constructed for multiple operating points are now used to compute the corresponding $K_{\rm ss}^i$ for various sensor modes of the \eqref{7}. The matrices $A|_{X_{{\rm{ss}}_i}}$, $C|_{X_{{\rm{ss}}_i}}$ and $K_{\rm ss}^i$ are finally stored in a look-up table and are used to construct multiple HKFs (MHKFs) as given by \eqref{6} for all the sensor modes.
Therefore, the HKF for the $j^{th}$ sensor mode at the $i^{th}$ operating point is designed as follows:
\begin{align}
&\hat{X}^{(i,j)}(k+1)-X_{\rm{OBEM}}(k+1)=A^i(\hat{X}^{(i,j)}(k)-X_{\text{OBEM}}(k))+K_{\rm ss}^i(Y(k)-\hat{Y}^{(i,j)}(k)),\notag\\
&\hat{Y}^{(i,j)}(k)= C^i( \hat{X}^{(i,j)}(k)-X_{\text{OBEM}}(k))+Y_{\text{OBEM}}(k)+b_{dj} a_{j}\delta_{j}(k),
\label{8}	
\end{align}
\noindent where $i=1,\ldots,L$, $j=1,\ldots,(q+1)$, $b_{dj}$ denotes the pre-determined sensor bias fault that can be different from the actual sensor fault $b_s$ that is injected into \eqref{7}, $a_j$ denotes the $q$-dimensional vector and is one of $q+1$ modes of $a$ which is the fault parameter vector. For the healthy mode or $j=1$, the fault parameter vector, $a$, is set to a zero vector whereas for $j= 2, \ldots, q+1$, $a_j$ has a unit value for the $(j-1)^{th}$ element and all the other elements are set to zero. Moreover, $A^i=A|_{X_{{\rm{ss}}_i}}$, $C^i=C|_{X_{{\rm{ss}}_i}}$ and $K_{\rm ss}^i$ are the previously stored state-space and Kalman gain matrices that depend on the $i^{th}$ operating point. Therefore, $L\times (q+1)$ HKFs are constructed covering the entire engine operating range corresponding to different sensor modes. It must be noted that the state-space matrices depend only on the operating points but the Kalman gain matrices depend on both the operating points and the noise covariance matrices that are the same for all the sensor fault modes, given that the OBEM does not take into account the effects of sensor faults. 

Although the operating range of a PWL model in the HKF scheme has been increased as compared to that of the one that uses the standard linear Kalman filters by replacing the steady-state values with the OBEM variables, still none of the PWL models are solely valid over the entire operating range of an engine. Therefore, corresponding to each PWL model one can associate a validity function that is based on its normalized weight as obtained by means of the Bayes formula. For this purpose, the residual vectors $\gamma^{(i,j)}$ and the covariance matrices $S^{(i,j)}$ that are generated by the multiple HKFs are used to compute the likelihood function $f^{(i,j)}$ for the $j^{th}$ sensor mode at the $i^{th}$ operating region as follows:
\begin{align}
&\gamma^{(i,j)}(k)=Y(k)-\hat{Y}^{(i,j)}(k), \hspace{2mm} S^{(i,j)}(k)=\text{cov}(\gamma^{(i,j)}(k)),\notag \\
&f^{(i,j)}(\gamma^{(i,j)}(k))=\frac{1}{(2\pi)^{q/2}\sqrt{\left|S^{(i,j)}(k)\right|}}\times\text{exp}[\frac{-1}{2}(\gamma^{(i,j)}(k))^{\rm T}(S^{(i,j)}(k))^{(-1)}(\gamma^{(i,j)}(k))],
\label{9}	
\end{align}
\noindent where it is assumed that the innovation sequence generated by the hybrid Kalman filter, $\gamma^{(i,j)}(k)$, is a Gaussian white noise process with zero mean and covariance matrix $S^{(i,j)}(k)$ that is calculated numerically. The normalized weights for the $j^{th}$ sensor mode are updated recursively by using the Bayes formula as follows:
\begin{align}
&w^{(i,j)}(k)=\frac{f^{(i,j)}(\gamma^{(i,j)}(k))w^{(i,j)}(k-1)}{\sum_{i=1}^{L}f^{(i,j)}(\gamma^{(i,j)}(k))w^{(i,j)}(k-1)}.
\label{10}	
\end{align}

The weights computed above should also remain outside a narrow bound to avoid becoming close to zero as:
\begin{align}
& \text{if} \; w^{(i,j)}(k)>\rho \; \text{then} \; w^{(i,j)}(k)= w^{(i,j)}(k), \notag \\
& \text{if} \; w^{(i,j)}(k)\leq  \rho   \; \text{then} \; w^{(i,j)}(k)=\rho,
\label{11}
\end{align}
\noindent where $\rho$ is a design parameter that is determined by trial and error and it invokes the PWL models that have  very small weights and it avoids them from being removed from the set of $L$ models. It is also useful for numerical robustness of the recursive weight algorithm \cite{Li2000}. 

Following the computation of the normalized weights, multiple time-varying models are now constructed that cover the entire engine operating range subject to various sensor modes. For each sensor mode, the corresponding model state-space matrices as well as the weighted innovation vector and the covariance matrix are obtained by using the PWL model state-space matrices and their associated normalized weights as follows:
\begin{alignat}{2}
&A_{c}^{j}(k)=\sum_{i=1}^{L}w^{(i,j)}(k)A^i, \qquad \qquad
C_{c}^{j}(k)=\sum_{i=1}^{L}w^{(i,j)}(k)C^i, \notag\\
&\gamma_{c}^{j}(k)=\sum_{i=1}^{L}w^{(i,j)}(k)\gamma^{(i,j)}(k),\qquad
S_{c}^{j}(k)=\sum_{i=1}^{L}(w^{(i,j)}(k))^2S^{(i,j)}(k),
\label{12}	
\end{alignat}
\noindent where $A_c^j(k)$ and $C_c^j(k)$ denote the weighted state-space matrices of a linear time-varying model associated with the $j^{th}$ sensor mode. Also, $\gamma_c^j(k)$ and $S_c^j(k)$ denote the weighted innovation vector and the covariance matrix of the $j^{th}$ sensor mode, respectively. The above procedure is now designated as the \emph{PWL models interpolation}. Consequently, $(q+1)$ weighted innovation vectors and covariance matrices that operate through out the entire engine operational regime are used in the next two sections to develop our proposed FDII scheme.

\section{SENSOR FDI VIA MULTIPLE-MODEL-BASED SCHEME}
\label{MM method}
In this section, the overall structure of our proposed MM-based FDI scheme is presented. It is assumed that the fault parameter vector can take on only one of $(q+1)$ sensor modes as $a_j$. Therefore, at each operating point there are $(q+1)$ PWL models; one for the healthy sensors scenario and $q$ corresponding to various faulty sensor scenarios, that have been designed and integrated with only one OBEM for constructing the multiple HKF (MHKF)-based scheme as formulated in \eqref{8}. The innovation vectors and the covariance matrices that are generated by the MHKFs are fused and weighted as given by \eqref{12}. Finally, there are $(q+1)$ weighted innovation vectors and covariance matrices that are used in the MM-based FDI scheme, where they are  operating under different healthy and faulty sensor scenarios during the entire engine operational regime.
\vspace{-3mm}
\subsection{Single Fault Detection and Isolation (FDI) Scheme}
In the MM-based approach \cite{Naderi2012, Meskin2013, Meskin2010}, the hypothesis conditional probability $P_j(k)$ is defined as the probability that the fault parameter $a$ assumes the mode $a_j$, $j = 1,\ldots, q+1$, conditioned on the observed measurement history up to the $k^{th}$ sample, that is: $P_j(k)=\text{Pr}[a=a_j|\mathcal{Y}(k)=\mathcal{Y}_k]$,
where $\mathcal{Y}(k)$ is the random vector measurement history with $Y(1), Y(2),\ldots, Y(k)$ partitions displaying the available measurements up to the $k^{th}$ sample time. Similarly, $\mathcal{Y}_k$ is the measurement history vector realization that has the partitions of $Y_1, Y_2,\ldots, Y_k$. Therefore, the conditional probability can be computed recursively as follows:
\begin{equation}
\hspace{-2mm}P_j(k)=\frac{f_{Y(k)|a,\mathcal{Y}(k-1)}(Y_j|a_j,\mathcal{Y}_{k-1})P_j(k-1)}{\sum_{\hbar=1}^{q+1}f_{Y(k)|a,\mathcal{Y}(k-1)}(Y_j|a_{\hbar},\mathcal{Y}_{k-1})P_{\hbar}(k-1)},
\label{14}	
\end{equation}

\noindent where $f_{Y(k)|a,\mathcal{Y}(k-1)}(Y_j|a_j,\mathcal{Y}_{k-1})$ denotes the Gaussian density function for the current measurement given by:
\begin{align}
&f_{Y(k)|a,\mathcal{Y}(k-1)}(Y_j|a_j,\mathcal{Y}_{k-1})=\frac{1}{(2\pi)^{q/2}\sqrt{\left|S_{c}^{j}(k)\right|}}\times \text{exp}[ \frac{-1}{2}(\gamma_{c}^{j}(k))^{\rm T}(S_{c}^{j}(k))^{(-1)}(\gamma_{c}^{j}(k))],
\label{15}	
\end{align}

\noindent where $\gamma_{c}^{j}(k)$ and $S_{c}^{j}(k)$ are given by  \eqref{12}.

\begin{figure*}
	\centering
	\includegraphics[width=\textwidth]{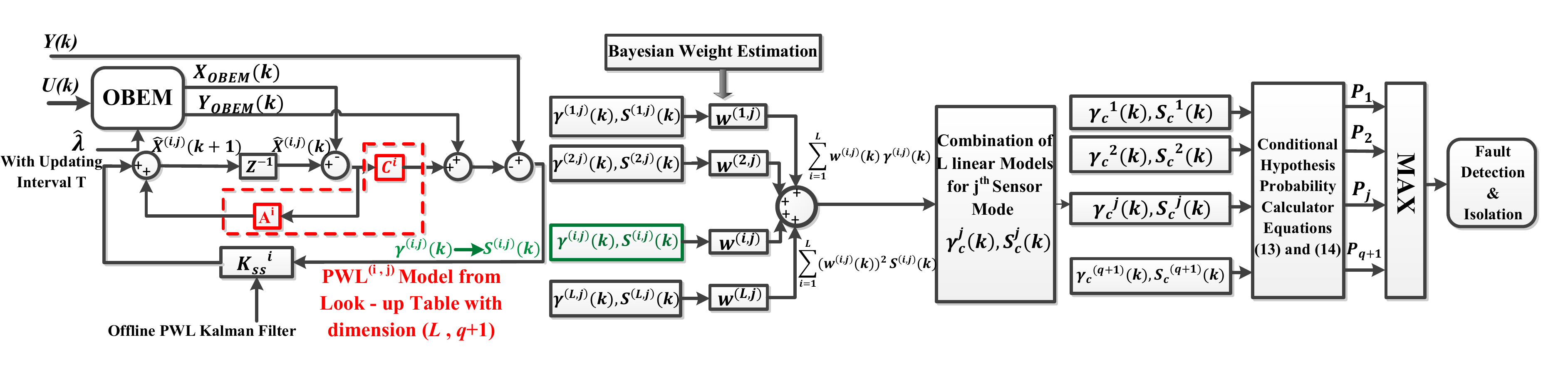}
	\vspace{-10mm}
	\caption{The MM-based FDI scheme based on the HKF corresponding to multiple operating regime.}
	\label{FDI_HKF}
	\vspace{-4mm}
\end{figure*}

If the $j^{th}$ sensor mode occurs, the probability associated with the $j^{th}$ model will be larger than that of the others since its corresponding innovation vector and also the determinant of the covariance matrix will be much smaller than those that are predicted by the other filters and which are mismatched with the assumed fault scenario. Hence, the condition of the system and the location of a single faulty sensor can be detected and isolated based on evaluating $P_j(k)$ and determining its maximum value. Consequently, our proposed MM-based approach is capable of detecting and isolating sensor faults. \figurename{ \ref{FDI_HKF}} shows the structure of the MM-based FDI scheme that employs the MHKFs for the entire engine operational regime. 

Our contribution here is the modification of our previously developed MM-based structure in \cite{Meskin2010} and \cite{Meskin2013} that utilizes only multiple standard linear Kalman filters and is designed for a single operating point into a general strategy that is applicable to the entire engine operational regime by means of the multiple HKF scheme and fusion and integration of the corresponding innovation vectors and covariance matrices.

As we will describe in more detail in Section \ref{SIMULATION RESULTS}, in this paper five sensor faults are considered. Therefore, the total number of modes corresponding to each operating point of the jet engine is six, where mode \#1 ($P_1$) corresponds to the healthy engine sensors and modes \#2 to \#6 ($P_2$ to $P_6$) correspond to a $3\%$ sensor bias fault injected into sensors measuring the compressor exit temperature ($P_2$) and pressure ($P_3$), shaft rotational speed ($P_4$), the turbine exit temperature ($P_5$) and pressure ($P_6$), respectively.
\vspace{-3mm}
\subsection{Concurrent Fault Detection and Isolation (FDI) Scheme}
\label{Conc}
For detection and isolation of two concurrent faults in the gas turbine engine, a hierarchical MM approach is proposed in \cite{Maybeck1999} and \cite{Meskin2013} as shown in \tablename{} \ref{Concurrent Fault Des}. In this scheme, it is assumed that the engine starts operating with healthy sensors when the first level of filters are active and the FDI scheme observes the sensors condition for occurrence of one of the five faulty modes. The active bank of filters operate based on the weighted innovation vectors and covariance matrices that are obtained in Section \ref{PWL}.
\begin{table} [!t]
	\caption{Operational modes corresponding to various possible two concurrent sensor faults.}
  \centering
  \footnotesize{
  \vspace{-1mm}
 \begin{tabular}{|c|c|c|c|c|c|c|} \hline
\multirow{2}{*} {Level} & \IEEEeqnarraymulticol{5}{t}{\hspace{6mm}Operational Modes} & \\ \cline{2-7}
& $\#1$ & $\#2$ & $\#3$ & $\#4$ & $\#5$ & $\#6$ \\ \hline
{First} & {Healthy} & $P_2$ & $P_3$ & $P_4$ & $P_5$ & $P_6$ \\ \hline
\multirow{10}{*} {Second} & $P_2$& $P_2$ & $P_2$ & $P_2$ & $P_2$ & $P_2$ \\   
&  & $P_2$ & $P_3$ & $P_4$ & $P_5$ & $P_6$ \\ \cline{2-7}
& $P_3$ & $P_3$ & $P_3$ & $P_3$ & $P_3$ & $P_3$   \\
&  & $P_2$ & $P_3$ & $P_4$ & $P_5$ & $P_6$ \\ \cline{2-7}
& $P_4$ & $P_4$ & $P_4$ & $P_4$ & $P_4$ & $P_4$  \\
&  & $P_2$ & $P_3$ & $P_4$ & $P_5$ & $P_6$ \\ \cline{2-7}
& $P_5$ & $P_5$ & $P_5$ & $P_5$ & $P_5$ & $P_5$  \\
&  & $P_2$ & $P_3$ & $P_4$ & $P_5$ & $P_6$ \\ \cline{2-7}
& $P_6$ & $P_6$ & $P_6$ & $P_6$ & $P_6$ & $P_6$   \\
& & $P_2$ & $P_3$ & $P_4$ & $P_5$ & $P_6$ \\ \hline
\end{tabular}}
\label{Concurrent Fault Des}
\vspace{-4mm}
\end{table}

Once the first fault is detected and isolated according to the maximum probability criterion, the FDI scheme will activate the second level of the filters as shown in \tablename{} \ref{Concurrent Fault Des} for detection of the second concurrent faults. In our hierarchical MM scheme, it is assumed that the sensor faults do not occur simultaneously and there exists always a minimum time interval between the occurrence of two sensor faults that are called concurrent faults. \tablename {} \ref{Concurrent Fault Des} indicates all the possible configurations for the second bank of filters that are considered in the simulation results in Section \ref{SIMULATION RESULTS}. The first filter in the second level always corresponds to the detected faulty mode in the first level. For example, if the first filter detects a $3\%$ bias in the compressor exit temperature sensor ($P_2$), then the first and the second filters in the second level respectively correspond to $P_2$ and the possibility of a larger bias fault within the compressor exit temperature sensor. Moreover, the third filter corresponds to the concurrent bias faults in the compressor exit temperature and pressure sensors ($P_2$ and $P_3$), the forth filter corresponds to $P_2$ and $P_4$, etc. This structure can easily be extended to the third and higher levels that correspond to occurrence of multiple concurrent sensor faults. It must be noted that when the new bank of filters is activated in the second level, the first bank will be disabled in order to avoid adding any unnecessary computational burden. Therefore, at any given time only six filters are operating on-line.

\textbf{Remark 1.} In this paper, the value of the pre-determined sensor bias fault severity, $b_{dj}$ as given by \eqref{8}, is considered to be the same for all the filters and only one level of fault severity is considered for the hierarchical MM scheme, although various pre-determined sensor bias faults can easily be incorporated into our strategy by correspondingly increasing the number of filters that are used in this scheme.
\section{SENSOR FAULT IDENTIFICATION/ESTIMATION}
\label{FID}
One of the important requirements of a general control system is the capability to integrate the FDI scheme with a fault identification or estimation module in order to estimate the severity of a fault that has occurred in different components of the system such as sensors and actuators. In this paper, a modified version of the generalized likelihood ratio (GLR) scheme is developed to estimate the severity of a sensor bias fault. The GLR was initially proposed by Willsky in \cite{Willsky1976} and was subsequently modified in \cite{Prakash2002, Prakash2005, Patwardhan2006} and \cite{Deshpande2008}. This method is capable of estimating the time, location and severity of an occurred fault using a selected threshold. However, in this paper the occurrence time and location of a fault have already been determined by means of our proposed MHKF-based approach as presented in Section \ref{MM method}. Hence, the fault severity will be estimated through the development of a modified version of the GLR scheme. The GLR method is a detection-estimation scheme in which the time of a fault is an \emph{a priori} information input to the estimation stage.
In this paper, the GLR method is simplified by removing the detection process through integrating the GLR method with our proposed MM-based FDI scheme, and hence there is no longer a need to select a threshold for our FDII scheme.

Let us assume that a bias fault has occurred in the $s^{th}$ sensor with the severity $b_s$ at the sample $k_{fs}$ and it is detected and isolated at the sample $k_{ds}$ when the $P_1(k)$ mode probability intersects with the $P_{(s+1)}(k)$ mode probability. Also, for $k\geq k_{ds}$, $P_{(s+1)}(k)$ has the highest value among the other mode probabilities as shown mathematically by $(s+1)={\rm{arg}}_j\;{\rm{max}}\; P_j(k)$. Therefore, the effects of the $s^{th}$ sensor fault still remain in the residual of the MHKF that is associated with the healthy mode. There are $L$ PWL models corresponding to the healthy mode that will be used subsequently to estimate the fault severity as follows:
\begin{align}
&\hat{X}^{(i,1)}(k+1)-X_{\rm{OBEM}}(k+1)=A^i(\hat{X}^{(i,1)}(k)-X_{\text{OBEM}}(k))+K_{ss}^i(Y(k)-\hat{Y}^{(i,1)}(k)), \notag\\
&\hat{Y}^{(i,1)}(k)= C^i(\hat{X}^{(i,1)}(k)-X_{\text{OBEM}}(k))+Y_{\text{OBEM}}(k),
\label{16}
\end{align}
\noindent in which the $j^{th}$ index and $b_{dj}$ in \eqref{8} are equal to one and zero, respectively, that are designated to the healthy sensor mode. Therefore, the residual vector of the MHKF that is designed for the healthy sensor mode at any subsequent time can be expressed as follows:
\begin{align}
& \gamma^{(i,1)}(k)=b_s G_s^i(k,k_{ds})z_s+v(k) \quad \text{for} \; k\geq k_{ds},
\label{17}
\end{align}
\noindent where $z_s$ is defined in \eqref{7} and $G_s^i(k,k_{ds})$ denotes the failure signature matrix that provides one with the information on the failure propagation through the filter, and which also depends on both $k_{ds}$ and the sample time $k$ at which the set of $L$ innovation vectors are computed for the MHKFs associated with the healthy mode. The signature matrices for a sensor bias fault can be computed by using the recursive relations for the healthy mode MHKFs as follows \cite{Prakash2002}:
\begin{align}
&G_s^i(k,k_{ds})=I-C^i A^iJ_s^i(k-1,k_{ds}),\notag
\end{align}
\begin{align}
&J_s^i(k,k_{ds})=A^iJ_s^i(k-1,k_{ds})+K_{ss}^i G_s^i(k,k_{ds}),
\label{18}
\end{align}
\noindent where $J_s^i(k,k_{ds})$ denotes the signature matrix for the state correction. The signature matrices $J_s^i(k,k_{ds})$ and $G_s^i(k,k_{ds})$ are defined separately for the $L$ operating points over a window of duration $[k_{ds},k_{ds}+\mathcal{N}]$, where $\mathcal{N}$ denotes the data samples window length. Also, it is assumed that $J_s^i(k_{ds}-1,k_{ds})=0$. The weighted signature matrices for a detected fault that cover the entire operating range of the gas turbine engine can be computed as follows:
\vspace{-2mm}
\begin{align}
& G_s(k,k_{ds})=\sum_{i=1}^L w^{(i,1)}(k)G_s^i(k,k_{ds}), \notag\\
& J_s(k,k_{ds})=\sum_{i=1}^L w^{(i,1)}(k)J_s^i(k,k_{ds}),
\label{19}
\end{align}
\noindent where $w^{(i,1)}(k)$ denotes the assigned weight to the $i^{th}$ PWL model that is designed for the healthy sensor mode. Therefore, the modified GLR test which consists of a maximum likelihood estimation (MLE) of $b_s$ when $k_{ds}$ is known is used. For performing the maximization process, normally the log likelihood function is chosen as the GLR criterion as follows:
\begin{align}
\label{20e}
\mathcal{J}=\sum_{k=k_{ds}}^{k_{ds}+\mathcal{N}}{\gamma_c^1}^T(k)({S_c^1}(k))^{(-1)}\gamma_c^1(k)&-\sum_{k=k_{ds}}^{k_{ds}+\mathcal{N}}[\gamma_c^1(k)-b_s G_s(k,k_{ds})z_s]^T({S_c^1}(k))^{(-1)}[\gamma_c^1(k)\notag\\
&-b_s G_s(k,k_{ds})z_s],
\end{align}

\noindent where $\gamma_c^1(k)$ and $S_c^1(k)$ denote the weighted innovation vector and covariance matrix for the healthy sensor mode that are computed from \eqref{12}. The maximization of the log likelihood function is associated with the minimization of the second summation in \eqref{20e}. It can easily be shown that the optimal solution to the MLE problem above is the estimated fault $\hat{b}_s=\frac{d_s}{c_s}$, that is the unbiased estimate of $b_s$ having the minimum variance as follows:
\begin{align}
& d_s=z_s^{\rm T}\sum_{k=k_{ds}}^{k_{ds}+\mathcal{N}}G_s^{\rm T}(k,k_{ds})({S_c^1}(k))^{(-1)}\gamma_c^1(k), \notag\\
& c_s=z_s^{\rm T}\sum_{k=k_{ds}}^{k_{ds}+\mathcal{N}}G_s^{\rm T}(k,k_{ds})({S_c^1}(k))^{(-1)} G_s(k,k_{ds})z_s.
\label{20}
\end{align}

The above procedure can also be employed for estimating the severities of multiple concurrent faults that have already been detected and isolated by means of our proposed hierarchical MM-based FDI approach.

\textbf{Remark 2.} In this paper, it is assumed that there is no  feedback and information sent from the fault identification module to the FDI scheme; otherwise, the FDI scheme needs to have a variable structure for updating the pre-determined sensor bias faults based on the estimated faults severities. 
\vspace{-2mm}
\section{SIMULATION RESULTS}
\label{SIMULATION RESULTS}
In this section, simulation results and performance evaluation of our proposed sensor FDII scheme as applied to several fault scenarios are presented for a nonlinear mathematical model of a commercial single spool jet engine previously developed in \cite{Naderi2012}. This model is generated based on the rotor, volume and heat transfer dynamic behavior and is also validated by the commercially available standard software \emph{GSP 10} \cite{GSP}. The responses corresponding to our mathematical model and the GSP match each other within an acceptable error tolerance (below $5\%$). For this study, our model is extended to the entire flight profile and is simulated in the SIMULINK for use as both the actual engine and the OBEM model. The actual engine operates at a given health condition subject to the effects of process and measurement noise signals, whereas the OBEM health parameters reference baselines are periodically updated to their recently estimated values that are assumed to be generated by an off-line health monitoring module as described in Section \ref{HKF}. 

Both the actual engine and the OBEM models operate in almost the same ambient conditions. The set of nonlinear state-space equations that are used for the engine model are given as follows:
\begin{align}
&\dot{P}_{{\rm CC}}=\frac{P_{{\rm CC}}}{T_{{\rm CC}}c_v \dot{m}_{{\rm CC}}}[(c_pT_{\rm C}\lambda_{\dot{m}_{\rm C}}\dot{m}_{\rm C}+\eta_{{\rm CC}}H_u\dot{m}_{\rm f}-c_pT_{{\rm CC}}\lambda_{\dot{m}_{\rm T}}\dot{m}_{\rm T})-c_v T_{{\rm CC}}(\lambda_{\dot{m}_{\rm C}}\dot{m}_{\rm C}+\dot{m}_{\rm f}\notag\\
&\qquad-\lambda_{\dot{m}_{\rm T}}\dot{m}_{\rm T})]+\frac{\gamma RT_{{\rm CC}}}{V_{{\rm CC}}}(\lambda_{\dot{m}_{\rm C}}\dot{m}_{\rm C}+\dot{m}_{\rm f}-\lambda_{\dot{m}_{\rm T}}\dot{m}_{\rm T}),\notag\\
&\dot{N}=\frac{\eta_{{\rm m}}\lambda_{\dot{m}_{\rm T}}\dot{m}_{\rm T}c_p(T_{{\rm CC}}-T_{\rm T})-\lambda_{\dot{m}_{\rm C}}\dot{m}_{\rm C}c_p(T_{\rm C}-T_{\rm d})}{JN(\frac{\pi}{30})^2}, \notag\\
&\dot{T}_{{\rm CC}}=\frac{1}{c_v m_{{\rm CC}}}[(c_pT_{\rm C}\lambda_{\dot{m}_{\rm C}}\dot{m}_{\rm C}+\eta_{{\rm CC}}H_u\dot{m}_{\rm f}-c_pT_{{\rm CC}}\lambda_{\dot{m}_{\rm T}}\dot{m}_{\rm T})-c_v T_{{\rm CC}}(\lambda_{\dot{m}_{\rm C}}\dot{m}_{\rm C}+\dot{m}_{\rm f}-\lambda_{\dot{m}_{\rm T}}\dot{m}_{\rm T})], \notag\\
&\dot{P}_{\rm T}=\frac{RT_{\rm M}}{V_{\rm M}}(\lambda_{\dot{m}_{\rm T}}\dot{m}_{\rm T}+\frac{\beta}{\beta+1}\lambda_{\dot{m}_{\rm C}}\dot{m}_{\rm C}-\dot{m}_{\rm n}),
\label{21}	
\end{align}

\noindent where $X=[P_{\rm CC},N,T_{\rm CC},P_{\rm T}]^{\rm T}$ denotes the state variable vector that includes the combustion chamber pressure $P_{\rm CC}$ and temperature $T_{\rm CC}$, the rotational speed $N$ and the turbine pressure $P_{\rm T}$. There is a single actuator that supplies the fuel flow ($u=\dot{m}_{\rm f}$) as well as five sensors measuring $Y=[T_{\rm C},P_{\rm C},N,T_{\rm T},P_{\rm T}]^{\rm T}$, where $T_{\rm C}=T_{\rm d} \left[ 1+\frac{1}{\lambda_{\eta_{\rm C}}\eta_{\rm C}}[(\frac{P_{\rm CC}}{P_{\rm d}})^{\frac{\gamma-1}{\gamma}}-1]\right]$ and $T_{\rm T}=T_{\rm CC}\left[1-\lambda_{\eta_{\rm T}}\eta_{\rm T}[1-(\frac{P_{\rm T}}{P_{\rm CC}})^{\frac{\gamma-1}{\gamma}}]\right]$ denote the  compressor and the turbine temperatures, respectively. The variables $P_{\rm C}$ and $P_{\rm T}$ denote the pressures of the compressor and turbine, respectively.
Moreover, $H=[\eta_{\rm C},\eta_{\rm T},\dot{m}_{\rm C},\dot{m}_{\rm T}]^{\rm T}$ denotes the health parameter vector, where $\eta_{\rm C}$ and $\eta_{\rm T}$ denote the compressor and the turbine efficiencies, and $\dot{m}_{\rm C}$ and $\dot{m}_{\rm T}$ denote their  mass flow rates, respectively. The health parameters are also multiplied by their corresponding health parameter degradation factors or the reference baselines $\lambda=[\lambda_{\eta_{\rm C}},\lambda_{\eta_{\rm T}},\lambda_{\dot{m}_{\rm C}},\lambda_{\dot{m}_{\rm T}}]^{\rm T}$. The sensors are affected by the Gaussian measurement noise with the standard deviations of $SD_{v}=[0.23,0.164,0.051,0.097,0.164]^{\rm T}$ corresponding to the percentages of output vector at the cruise condition as defined in \cite{Naderi2012}. More details on other variables and constants in \eqref{21} are provided in \cite{Naderi2012}.

\begin{figure*}
	\centering	
	\includegraphics[width=\textwidth]{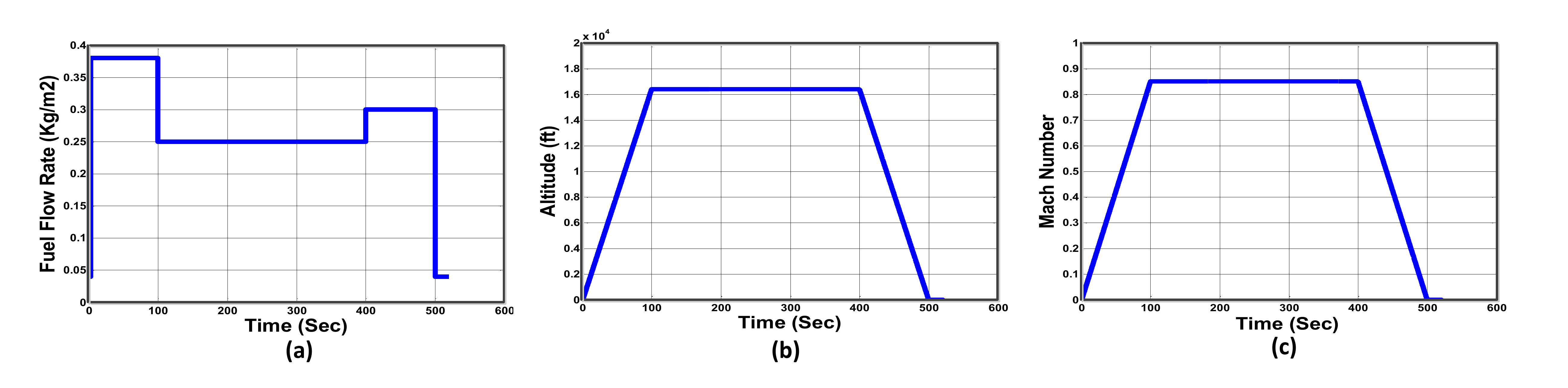}
	\vspace{-6mm}
	\caption{Profiles of (a) the fuel flow rate, (b) altitude, and (c) the Mach number during a flight mission.}		
	\label{Flight_Mission}
	\vspace{-2mm}
\end{figure*}

The flight condition is defined by two environmental variables, namely the altitude and the Mach number. The ambient temperature and pressure can be computed according to $T_{\text{amb}}=T_s-\frac{6.5Alt}{1000}$ and $P_{\text{amb}}=P_s \text{exp}(\frac{-gMAlt}{288R})$, where $T_s = 288^ {\;\circ} {\rm K}$ and $P_s = 1.01325$ bar are set to the standard condition, $R = 8.31447$ is the gas constant, $g$ is the gravitational acceleration, and $M$ and $Alt$ denote the Mach number and the altitude, respectively. Moreover, the ambient variables are affected by the Gaussian process noise with the standard deviations of $SD_{\zeta}=[0.01,0.01]^{\rm T}$ corresponding to the percentages of standard conditions of ambient variables. The noise free values of the ambient parameters are applied to the OBEM model. Also, the same control input is applied to both the actual engine and the OBEM models.
The system is simulated for 520 sec with the sampling rate of 0.01 sec. The profiles of the altitude, Mach number and the fuel flow rate are shown in \figurename{}  \ref{Flight_Mission}.

According to the overall structure of our proposed scheme, as shown in \figurename {} \ref{FDI_HKF}, six PWL models are constructed $(q=5)$ corresponding to each operating point for the five faulty sensor modes as well as the one healthy mode with the pre-determined sensor bias faults set to 3\% of the engine steady-state outputs under the cruise condition. The states and sensor outputs are estimated by means of our proposed MHKFs. Finally, six weighted matrices are calculated according to \eqref{12} and are applied in the MM-based scheme for detecting and isolating the sensor bias faults that occur at different points of the flight profile.

For our simulations, the measurement and process noise covariance matrices are set to $0.01I$ and $0.1I$, respectively, where $I$ denotes an identity matrix. In this paper, the FDII scheme is implemented for the entire flight profile including the climbing, cruise and landing modes. Therefore, we assume and consider five operating points handling the entire flight profile $(L=5)$. This is the \textit{minimum} number of the operating points that are obtained for each faulty mode that prevents the occurrence of a false alarm in the range of the applied health degradations. The number of the operating points depends on the dynamics of the engine as well as the range of the applied fuel flow rate and the environmental parameters. We simulated our proposed FDII scheme with different number of operating points and also derived a confusion matrix for each case in order to analyze the false alarm rates and to decide on the minimum required number of the operating points that does not lead to any false alarms for the injected health degradations. These details are not included here due to space limitations.

\tablename {} \ref{tbl:PWL} shows the corresponding fuel flow rates as well as the flight conditions corresponding to all the applied PWL models that are designed for each sensor mode. In order to efficiently track the variations of the system input during the climbing and landing conditions, two operating points are selected for each of these two modes; whereas only one operating point is selected for the cruise condition given the presence of a  constant input.

During the fault detection and isolation process, a mode probability $P_j(k)$ is generated for each weighted model using \eqref{14} and by determining the maximum $P_j(k), \; j=1,\ldots, q+1$, the sensor fault is detected and isolated.
\begin{table}
	\caption{The operating point specifications corresponding to the designed PWL models, where $\dot{m}_{\rm f}$ and $Alt$ are measured in ${\rm Kg/m^2}$ and ${\rm ft}$, respectively.}
  \centering
  \small
 \begin{tabular}{|c|c|c|c|} \hline
 {Models Corresponding} \multirow{2}{*} & $\dot{m}_{\rm f}$ & $M$ & $Alt$ \\
 {to Flight Conditions} &  & & \\\hline
 \text{Operating Point 1 (Climbing)} & $0.38$ & $0.2109$ & $4070.538$ \\ \hline
 \text{Operating Point 2 (Climbing)} & $0.38$ & $0.6585$ & $12708.33$ \\ \hline
 \text{Operating Point 3 (Cruise)} & $0.25$ & $0.85$ & $16404.2$ \\ \hline
 \text{Operating Point 4 (Landing)} & $0.3$ & $0.5402$ & $10424.87$\\ \hline
 \text{Operating Point 5 (Landing)} & $0.3$ & $0.1203$ & $2322.835$ \\ \hline
 \end{tabular}
\label{tbl:PWL}
\vspace{-4mm}
\end{table}

As described in Section \ref{Introduction}, the health parameter degradations due to the engine aging is one of the non-fault related factors that should be considered for a reliable and an accurate FDII strategy. Therefore, the performance of the FDII strategy is evaluated in presence of compressor or turbine health parameters degradations from their healthy reference baselines.
Generally, there is a difference between the actual and the estimated health parameter reference baselines due to  off-line health monitoring estimation errors and also due to effects of noise and disturbances. Therefore, the robustness of our proposed FDII scheme will also be evaluated subsequently in this section with respect to the percentage of the reference baselines estimation errors.
\vspace{-3mm}
\subsection{\underline{Case 1: False Alarms Evaluation}}
\label{False Alarm}
\vspace{-1mm}
Many factors such as (i) the dynamic mismatch between the OBEM model and the actual engine model, (ii) the large estimation errors for an off-line health monitoring module, and (iii) the process and measurement noise, may lead to false alarm flags. In order to evaluate the reliability and the efficiency of our proposed FDII scheme in terms of avoiding false alarms, our proposed scheme is simulated over the entire flight profile as shown in \figurename {} \ref{Flight_Mission}.
\begin {table*}  
	\caption{Fault detection time (FDT) corresponding to the maximum tolerable percentage of the RBEE $(\overline{\Delta \lambda})$ at different stages of the flight profile.}	
	\label{FDT}
	\centering
	\small
	\begin{tabular}{|l|l l l|c|c|c|} \hline
		\multirow{2}{*}{Faulty Scenario} & \IEEEeqnarraymulticol{2}{t}	{\multirow{2}{*}{\hspace{20mm}Maximum Tolerable RBEE\%}} & & \IEEEeqnarraymulticol{2}{t} {\hspace{15mm} Sensor FDT (sec) } & \\ \cline{5-7}
		 & & & & $k_f = 50$ & $k_f = 250$	& $k_f=450$ \\ \hline
	  \multirow{2}{*}{Fault on $T_{\rm C}$} & Compressor & $\overline{\Delta \lambda}_{\dot{m}_{\rm C}}=2.52$ & $\overline{\Delta \lambda}_{{\eta}_{\rm C}}=2.52$ & $3.7$ & $4.1$ & $6.3$ \\ & Turbine & $\overline{\Delta \lambda}_{\dot{m}_{\rm T}}=2.52$ & $\overline{\Delta \lambda}_{{\eta}_{\rm T}}=1.51$ & $3.5$ & $5.1$ & $5.9$ \\ \hline
	  \multirow{2}{*}{Fault on $P_{\rm C}$} & Compressor & $\overline{\Delta \lambda}_{\dot{m}_{\rm C}}=1.01$ & $\overline{\Delta \lambda}_{{\eta}_{\rm C}}=1.01$ & $7.8$ & $5.9$ & $8$ \\ & Turbine & $\overline{\Delta \lambda}_{\dot{m}_{\rm T}}=1.01$ & $\overline{\Delta \lambda}_{{\eta}_{\rm T}}=0.30$ & $3.4$ & $2.7$ & $2.5$ \\ \hline	
	  \multirow{2}{*}{Fault on $N$} & Compressor & $\overline{\Delta \lambda}_{\dot{m}_{\rm C}}=1.21$ & $\overline{\Delta \lambda}_{{\eta}_{\rm C}}=2.02$ & $6.5$ & $2.6$ & $3.5$ \\ & Turbine & $\overline{\Delta \lambda}_{\dot{m}_{\rm T}}=1.01$ & $\overline{\Delta \lambda}_{{\eta}_{\rm T}}=0.81$ & $6.5$ & $2.5$ & $2.9$ \\ \hline
	  \multirow{2}{*}{Fault on $T_{\rm T}$} & Compressor & $\overline{\Delta \lambda}_{\dot{m}_{\rm C}}=0.2$ & $\overline{\Delta \lambda}_{{\eta}_{\rm C}}=0.2$ & $7.8$ & $3$ & $4.7$ \\ & Turbine & $\overline{\Delta \lambda}_{\dot{m}_{\rm T}}=0.1$ & $\overline{\Delta \lambda}_{{\eta}_{\rm T}}=0.1$ & $6$ & $2.2$ & $2.3$ \\ \hline
	  \multirow{2}{*}{Fault on $P_{\rm T}$} & Compressor & $\overline{\Delta \lambda}_{\dot{m}_{\rm C}}=0.3$ & $\overline{\Delta \lambda}_{{\eta}_C}=0.3$ & $7$ & $2.6$ & $3.4$ \\ & Turbine & $\overline{\Delta \lambda}_{\dot{m}_{\rm T}}=0.3$ & $\overline{\Delta \lambda}_{{\eta}_{\rm T}}=0.1$ & $8$ & $2.2$ & $2.2$ \\ \hline	 	
	\end{tabular}
\end{table*}
The estimated reference baselines $(\hat{\lambda})$ may differ form the actual degraded engine $(\lambda)$ and this can play a source of uncertainty in our FDII scheme in addition to the process and measurement noise. It should be noted that the dynamic mismatch between the actual engine and the OBEM models is also another source of uncertainty, but this is not investigated in this paper. Therefore, the robustness of the FDII scheme against the percentage of the reference baselines estimation error RBEE (${\rm RBEE\%}=100\times \frac{\Delta \lambda}{\lambda}$, where $\Delta \lambda = |\lambda-\hat{\lambda}|$ and $\lambda \in [0,1]$) is investigated in this section.

Based on our simulation studies for the healthy sensors scenario, the FDII algorithm is robust to the maximum percentage of reference baselines estimation error of 3\% for the compressor health parameters and a maximum of 2\% for the turbine health parameters. Our proposed FDII algorithm has declared no false alarms in the range of the above RBEEs, however the possibility of false alarms will increase if the RBEEs are increased beyond these upper limits.
In case that one is within these limits, the mode probability of the healthy sensors is near one whereas the other probabilities corresponding to the faulty sensor scenarios are all almost zero during the entire flight profile. The values for the probabilities depend on the design parameter $\rho$. In general the MM-based FDI is independent of $\rho$ as long as it is selected as sufficiently small.

\subsection{\underline{Case 2: A 3\% Sensor Bias Fault Detection and Isolation}}
In this section, the performance of the sensor FDI scheme with respect to the fault detection time and the robustness of the algorithm with respect to the percentage of reference baselines estimation error (RBEE) are evaluated. This evaluation is performed during the entire flight profile which lasts for 520 sec, when a pre-determined bias fault with the severity of 3\% of the engine steady-state output values under cruise condition occurs for a single sensor.
\tablename{} \ref{FDT} shows the fault detection times for each single sensor fault scenario at different stages of the flight profile including the climbing, cruise and landing modes. It also shows the maximum tolerable percentages of the RBEE of engine health parameters that is denoted by $\overline{\Delta \lambda}$. It is assumed that $\lambda=0.99$. The limits are obtained through preserving two performance requirements for our proposed FDI scheme, namely: 1) the sensor fault detection time (FDT) should be less than 8 sec, and 2) no false alarms should be generated. This implies that our proposed FDI scheme can tolerate the reference baselines estimation error as long as the above performance requirements are fulfilled.

\begin{figure*}
   \centering
      \includegraphics[width=0.97\textwidth]{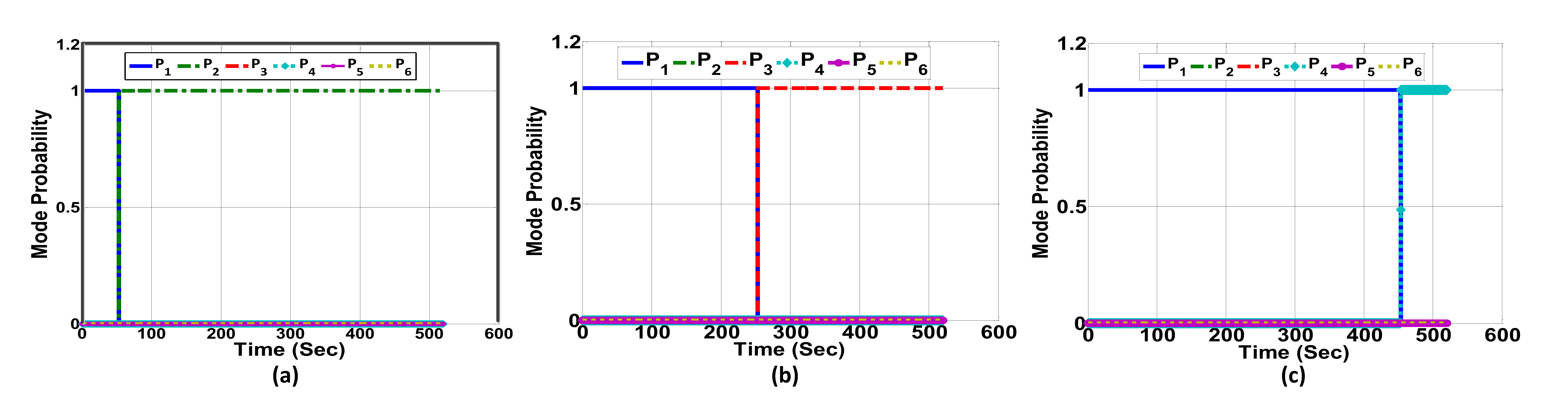}
      \vspace{-2mm}
      \caption{Mode probabilities for 3\% bias fault applied at (a) $k_f=50$ sec to the $T_{\rm C}$ sensor, (b) $k_f=250$ sec to the $P_{\rm C}$ sensor, and (c) $k_f=450$ sec to the $N$ sensor in presence of the obtained maximum tolerable percentage of the estimation errors in \tablename {} \ref{FDT} for the compressor health parameter reference baselines.}
      \label{Faulty_Prob}
      \vspace{-2mm}
\end{figure*}

In addition, \figurename{} \ref{Faulty_Prob} depicts the mode probabilities for three selected fault scenarios when the bias fault occurs at different instants of the flight profile. The value of the injected fault is set to 3\% of the engine steady-state output values, while the percentage of the estimation errors for the compressor health parameters reference baselines (RBEE) are set at their maximum tolerable levels that are indicated in \tablename{} \ref{FDT}.

By comparing the results in \tablename{} \ref{FDT}, it can be concluded that the sensor fault detection times during the cruise mode are much less than that of the other flight modes since there is less variation of thrust and ambient conditions. In spite of a large input and ambient condition variations during the climbing and the landing modes, it is still possible to detect a sensor fault by applying our proposed FDI scheme. In order to show this capability, the sensor faults occur at $k_f = 50$ sec during the climbing mode, at $k_f = 250$ sec during the cruise mode and  at $k_f = 450$ sec during the landing mode. The fault detection times are all indicated in \tablename {} \ref{FDT} for various fault scenarios.

To quantify the effectiveness and reliability of our proposed FDI scheme in presence of the mismatch between the OBEM model and the actual engine, confusion matrices are obtained for both the healthy and the faulty scenarios subject to different  health parameter degradations and  measurement noise. 
The rows in the confusion matrix are the fault conditions and the columns are the actual isolated faults. The element in the $i^{th}$ row and $j^{th}$ column (${\rm CM}_{ij}$) shows the rate that fault $j$ is isolated when fault $i$ occurs. Ideally, the confusion matrix should be a purely diagonal matrix. To obtain the confusion matrix for our sensor FDI scheme, 50 Monte Carlo simulations are performed in which the percentage of the RBEE exceeds the maximum tolerable limit that is reported in \tablename {} \ref{FDT}. Tables \ref{CM1} and \ref{CM2} show the confusion matrices for the healthy and $3\%$ faulty scenarios as applied to the gas turbine engine sensors during the cruise mode subjected to $3\%$ RBEE of both the compressor and the turbine health parameters, respectively.
\begin{table}
 \caption{The confusion matrix for $3\%$ RBEE of the compressor health parameters.}
 \label{CM1}
  \centering
  \footnotesize
  \vspace{-1mm}
 \begin{tabular}{|c|c|c|c|c|c|c|} \hline
              & $T_{\rm C}$ & $P_{\rm C}$ & $N$ & $T_{\rm T}$ & $P_{\rm T}$ & \text{No Fault} \\ \hline
 $T_{\rm C}$  & $50$  & $0$ & $0$ & $0$ & $0$ & $0$ \\ \hline
 $P_{\rm C}$  & $1$  & $49$ & $0$ & $0$ & $0$ & $0$ \\ \hline
 $N$          & $0$  & $0$ & $50$ & $0$ & $0$ & $0$ \\ \hline
 $T_{\rm T}$  & $12$  & $0$ & $0$ & $27$ & $10$ & $1$ \\ \hline
 $P_{\rm T}$  & $17$  & $0$ & $0$ & $0$ & $29$ & $4$ \\ \hline
 \text{No Fault} & $0$  & $0$ & $0$ & $0$ & $0$ & $50$ \\ \hline
 \end{tabular}
 \vspace{-2mm}
\end{table}

\begin{table}
 \caption{The confusion matrix for $3\%$ RBEE of the turbine health parameters.}
 \label{CM2}
  \centering
  \footnotesize
  \vspace{-1mm}
 \begin{tabular}{|c|c|c|c|c|c|c|} \hline
              & $T_{\rm C}$ & $P_{\rm C}$ & $N$ & $T_{\rm T}$ & $P_{\rm T}$ & \text{No Fault} \\ \hline
 $T_{\rm C}$  & $50$  & $0$ & $0$ & $0$ & $0$ & $0$ \\ \hline
 $P_{\rm C}$  & $0$  & $50$ & $0$ & $0$ & $0$ & $0$ \\ \hline
 $N$          & $1$  & $0$ & $48$ & $0$ & $0$ & $1$ \\ \hline
 $T_{\rm T}$  & $2$  & $0$ & $0$ & $41$ & $0$ & $7$ \\ \hline
 $P_{\rm T}$  & $0$  & $1$ & $0$ & $0$ & $44$ & $5$ \\ \hline
 \text{No Fault} & $0$  & $1$ & $0$ & $0$ & $0$ & $49$ \\ \hline
 \end{tabular}
 \vspace{-2mm}
\end{table}

In addition to the percentage of the RBEE, the measurement noise is another factor that can affect the performance of our proposed FDI scheme. Consequently, \tablename {} \ref{CM3} depicts  the confusion matrix that is obtained for investigating the effects of measurement noise signals, when their original standard deviations $(SD_v)$ are multiplied by a factor of $20$.

According to our simulations, the initial impact of any increase in either the standard deviation of the noise measurements or the percentage of the RBEE is to delay the fault detection time. Moreover, the occurrence of false alarms and incorrect fault detections are the other consequences due to increases in the above uncertainty sources. Different performance indices can be defined to quantify the robustness of our proposed sensor FDI algorithm with respect to the levels of uncertainty sources \cite{Fawcett2006}, namely:
\begin{align}
&\text{FPR}=\frac{\sum_{j=1}^5 {\rm CM}_{6j}}{\sum_{j=1}^6 {\rm CM}_{6j}}, \quad  \text{ACC}=\frac{\sum_{i=1}^6 {\rm CM}_{ii}}{\sum_{j=1}^6 \sum_{i=1}^6 {\rm CM}_{ij}}, \quad \text{IFDR}=\frac{\sum_{j=1}^5 \sum_{i=1}^5 {\rm CM}_{ij}(i\neq j)}{\sum_{j=1}^6 \sum_{i=1}^5 {\rm CM}_{ij}},
\label{25}
\end{align}

\noindent where FPR, ACC and IFDR denote the false positive (false alarm) rate, accuracy and incorrect fault detection rate, respectively, that are  calculated in \tablename {} \ref{CM_Ind} to investigate the effects of various uncertainty sources.
\begin{table}
 \caption{The confusion matrix for the measurement noise signals where the SDs are multiplied by a factor of $20$.}
 \label{CM3}
  \centering
  \footnotesize
\vspace{-1mm}
 \begin{tabular}{|c|c|c|c|c|c|c|} \hline
              & $T_{\rm C}$ & $P_{\rm C}$ & $N$ & $T_{\rm T}$ & $P_{\rm T}$ & \text{No Fault} \\ \hline
 $T_{\rm C}$  & $48$  & $0$ & $0$ & $0$ & $0$ & $2$ \\ \hline
 $P_{\rm C}$  & $0$  & $46$ & $0$ & $0$ & $0$ & $4$ \\ \hline
 $N$          & $0$  & $0$ & $49$ & $0$ & $0$ & $1$ \\ \hline
 $T_{\rm T}$  & $0$  & $0$ & $0$ & $48$ & $0$ & $2$ \\ \hline
 $P_{\rm T}$  & $0$  & $0$ & $0$ & $1$ & $47$ & $2$ \\ \hline
 \text{No Fault} & $0$  & $0$ & $0$ & $0$ & $1$ & $49$ \\ \hline
 \end{tabular}
 \vspace{-1mm}
\end{table}

\begin{table}
 \caption{Sensor FDI algorithm performance indices corresponding to different levels of uncertainties.}
 \label{CM_Ind}
  \centering
  \small
  \begin{tabular}{|c|c|c|c|} \hline
  \text{Scenarios} & FPR & ACC & IFDR \\ \hline
  $3\%$ \text{Compressor RBEE} & $0$ & $0.85$ & $0.16$\\ \hline
  $4\%$ \text{Compressor RBEE} & $0.12$ & $0.66$ & $0.376$\\ \hline
  $3\%$ \text{Turbine RBEE} & $0.02$ & $0.94$ & $0.016$\\ \hline
  $4\%$ \text{Turbine RBEE} & $0.04$ & $0.91$ & $0.076$\\ \hline
  \text{Noise SD}$\times 20$ & $0.02$ & $0.956$ & $0.0004$\\ \hline
  \text{Noise SD}$\times 25$ & $0.02$ & $0.91$ & $0.012$\\ \hline
 \end{tabular}
 \vspace{-3mm}
\end{table}
According to \tablename {} \ref{CM_Ind}, increasing the levels of the uncertainty sources results in decrease of ACC, but increase in FPR and IFDR. Notwithstanding these observations, our proposed sensor FDI scheme still works sufficiently robust against a high level of measurement noise as well as discrepancies between the OBEM model and the actual engine health parameters reference baselines.
\vspace{-2mm}
\subsection{\ul{Case 3: Sensor Fault Detection and Isolation for Different Fault Severities}}
In real applications, there is no guarantee that the sensor bias fault severity always matches the 3\% pre-determined fault for which the MM structure is designed. Therefore, it is essential to investigate the performance of our proposed MM-based FDI scheme for an applied sensor bias fault having different severities starting from the minimum detectable bias.
\tablename{} \ref{fault_severity} shows the average detection times for all the faulty modes as a function of the fault severities, when the applied faults occur at different stages of the flight profile with also the maximum tolerable RBEE\% for the compressor health parameters that are indicated in \tablename{} \ref{FDT}.
It can be observed from \tablename{} \ref{fault_severity} that the higher the fault severity with respect to the pre-determined $3\%$ fault, the later the detection time, given that the actual fault becomes further different from the pre-determined bias fault.
Note that the minimum detectable sensor bias fault is 2\% that requires larger time to be detected as compared to the higher fault severities, especially during the climbing and landing flight modes.

\begin{table}
 \caption{The average sensor fault detection times for all fault modes as a function of the fault severity at different stages of the flight profile, under the maximum tolerable RBEE\% for the compressor health parameters.}
 \label{fault_severity}
  \centering
  \small
  \vspace{-1mm}
 \begin{tabular}{|c|c|c|c|c|c|} \hline
 \text{Fault Time} & $2\%$ & $3\%$ & $4\%$ & $5\%$ & $6\%$ \\ \hline
 $k_f=50$  sec & $14.18$ & $6.56$ & $6.74$ & $7.42$ & $8.6$ \\ \hline
 $k_f=250$  \text{sec} & $8.34$ & $3.64$ & $3.98$ & $4.76$ & $5.7$ \\ \hline
 $k_f=450$  \text{sec} & $17.66$ & $5.18$ & $5.32$ & $6.66$ & $7.52$ \\ \hline
 \end{tabular}
 \vspace{-3mm}
\end{table}

It should be pointed out that our proposed MM-based structure is designed for a $3\%$ sensor bias fault and  is not capable of detecting and isolating sensor bias faults that have  far greater severities. For this purpose, \tablename{} \ref{MaxFault} shows the maximum detectable sensor bias faults by using our proposed FDI scheme by also considering the maximum tolerable RBEE\% for the compressor and the turbine health parameters. The reported detectable fault severities in \tablename{} \ref{MaxFault} correspond to the average fault severities for all the flight modes. However, if an applied sensor bias fault increases beyond the corresponding maximum detectable fault as indicated in \tablename{} \ref{MaxFault}, an incorrect fault may be detected. Therefore, it is recommended that one incorporates more models within the MM-based structure corresponding to higher pre-determined sensor faults to become capable of detecting and isolating faults with higher magnitudes in shorter durations of time.
\begin{table}
 \caption{The average of the maximum detectable sensor bias fault (as percentage of the engine steady-state outputs) over the entire flight profile by using our FDI scheme when the corresponding maximum tolerable RBEE\% reported in \tablename {} \ref{FDT} are applied to the compressor and turbine health parameters.}
 \label{MaxFault}
  \centering
  \small
  \vspace{-1mm}
 \begin{tabular}{|c|c|c|} \hline
 \text{Fault Modes} &  \multirow{2}{*}\text{Max. Tolerable} &  \multirow{2}{*}\text{Max. Tolerable} \\
 & \text{Compressor RBEE\%} & \text{Turbine RBEE\%}\\\hline
 $T_{\rm C}$ & $25\%$ & $70\%$ \\\hline
 $P_{\rm C}$ & $20\%$ & $50\%$ \\\hline
 $N$	& $12\%$ & $100\%$ \\\hline
 $T_{\rm T}$ & $10\%$ & $14\%$ \\\hline
 $P_{\rm T}$ & $70\%$ & $75\%$ \\\hline
 \end{tabular}
 \vspace{-3mm}
\end{table}
\vspace{-5mm}
\subsection{\underline{Case 4: Sensor Fault Severity Identification}}
One of the advantages of our proposed MM-based FDI scheme is its capability in providing  information on the occurred fault severity. However, this method is not precise for fault severity estimation, since it provides one with only information on the level of an occurred fault instead of the exact fault severity. Therefore, the modified GLR method introduced in Section \ref{FID} is integrated with our MM-based FDI scheme to estimate the sensor bias fault magnitude and severity.
In order to investigate the performance of our proposed fault severity estimation method, the notion of the percentage of weighted mean square normalized residual errors (WMSNE) are obtained for various detected fault scenarios as follows:
\begin{align}
\text{WMSNE}\%&=\notag\\
&\hspace{-10mm}\frac{1}{L}\sum_{i=1}^L\frac{\sum_{k=k_{ds}}^\mathcal{K} w^{(i,s+1)}(k)(\frac{(Y(k)-\hat{Y}^{(i,s+1)}(k))}{Y(k)})^2}{\sum_{k=k_{ds}}^\kappa w^{(i,s+1)}(k)}\times 100, \notag
\end{align}
\begin{align}
\hat{Y}^{(i,s+1)}(k)=C^i(\hat{X}^{(i,s+1)}(k)-X_{\text{OBEM}}(k))+Y_{\text{OBEM}}(k)+\hat{b}_s z_s\delta(k-k_{ds}),\quad s=1,\ldots, q
\label{26}
\end{align}

\noindent where $\mathcal{K}$ is the total simulation samples and $\hat{Y}^{(i,s+1)}(k)$ is the numerically constructed output associated with the $(s+1)^{th}$ operational mode corresponding to the $s^{th}$ fault, in which the fault vector effect is generated based on the estimated fault magnitude $\hat{b}_s$ and the fault occurrence detected time $k_{ds}$. It must be noted that there is no feedback and information sent from the fault estimation module to  the filters for  updating their pre-determined bias faults. \tablename{} \ref{WMSE} shows the average percentage of the WMSNE that is measured for different stages of the flight profile as a function of the bias fault severity as applied to various sensors.
\begin{table}[!h]
 \caption{The average WMSNE\% for all flight modes as a function of the bias fault severity applied to various sensors without considering any estimation error for the health parameter reference baselines.}
 \label{WMSE}
  \centering
  \small
  \vspace{-1mm}
 \begin{tabular}{|c|c|c|c|c|c|} \hline
 \text{Fault} & $2\%$ & $3\%$ & $4\%$ & $5\%$ & $6\%$ \\ \hline
 $T_{\rm C}$  & $0.0085$ & $0.0067$ & $0.0073$ & $0.0124$ & $0.0195$ \\ \hline
 $P_{\rm C}$  & $0.1360$ & $0.1106$ & $0.1293$ & $0.1689$ & $0.2537$ \\ \hline
 $N$          & $0.0111$ & $0.0109$ & $0.0107$ & $0.0105$ & $0.0104$ \\ \hline
 $T_{\rm T}$  & $0.0137$ & $0.0066$ & $0.0128$ & $0.0312$ & $0.0944$ \\ \hline
 $P_{\rm T}$  & $0.1313$ & $0.1269$ & $0.1359$ & $0.1503$ & $0.1615$ \\ \hline
  \end{tabular}
\end{table}

According to \tablename{} \ref{WMSE}, the average percentage of WMSNE is less than $0.5\%$ for various faults that are limited within the operational range of our FDI scheme as was defined in the preceding case. It can also be observed that the percentage of the WMSNE is lower for the cruise flight mode as compared to the climbing and landing modes due to the high variations of the engine thrust.
\vspace{-2mm}
\subsection{\ul{Case 5: Concurrent Fault Detection, Isolation and Identification}}
\begin{figure}
   \centering
   \vspace{-3mm}
   \includegraphics[width=0.97\columnwidth]{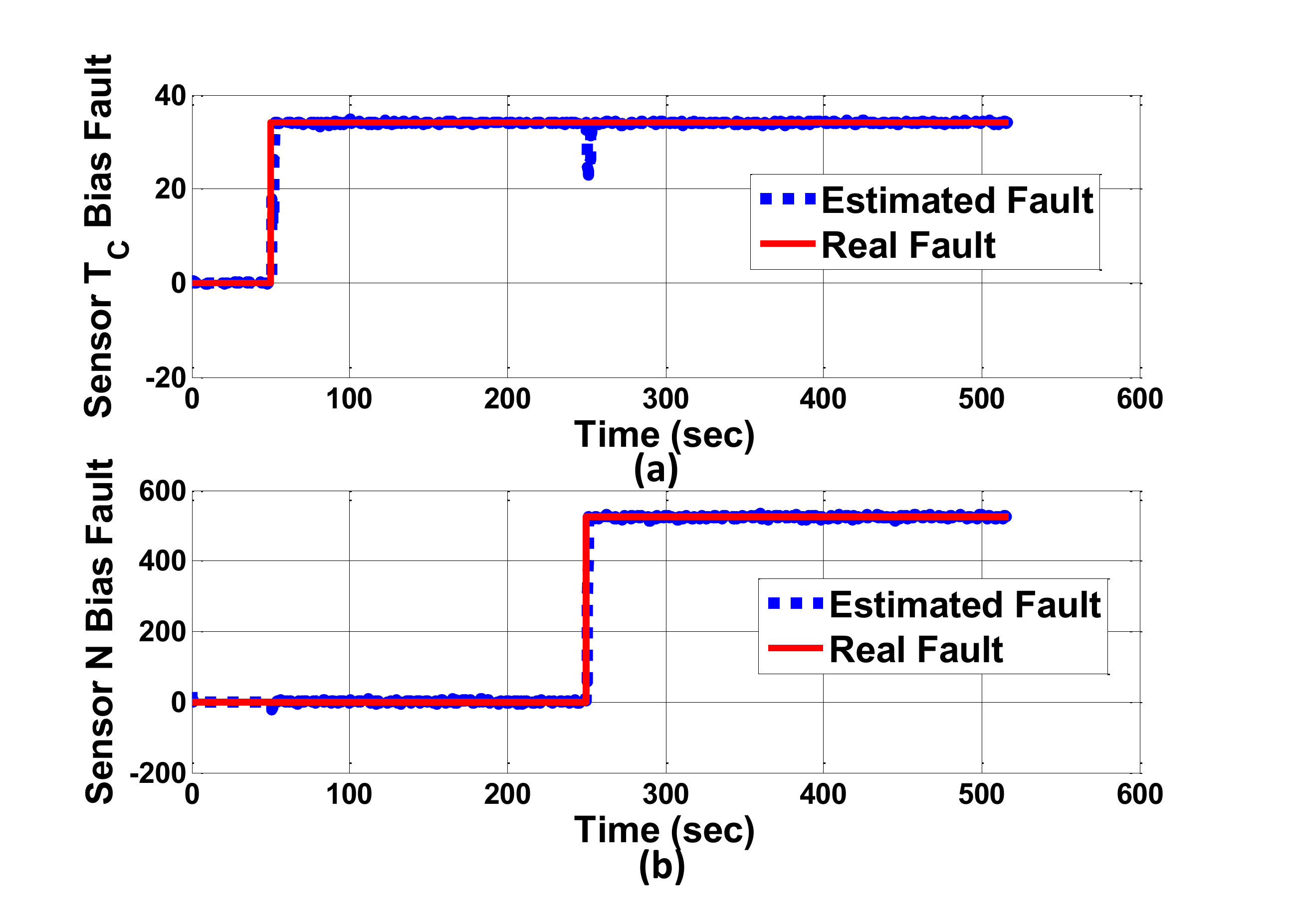}
   \vspace{-3mm}
   \caption{Estimated two concurrent sensor fault severities (a) a $6\%$ bias fault that is applied to the $T_{\rm C}$ sensor at $k_f=50$ sec, and (b) a $5\%$ bias fault that is applied to the $N$ sensor at $k_f=250$ sec without considering any health parameter degradations.}
   \label{FID_Conc1}
   \vspace{-4mm}
\end{figure}

In previous simulations only a single sensor fault is applied to the gas turbine engine. In this case study, our proposed FDII scheme is used for concurrent fault scenarios that occur at different stages of the flight profile. For this purpose, the hierarchical scheme described in Section \ref{Conc} is used to diagnose the concurrent sensor faults. The fault detection times and the percentages of the WMSNE as given by \eqref{26} are obtained for two selected  scenarios with concurrent sensor faults. The first scenario simulates the effects of a $6\%$ bias fault applied to the $T_{\rm C}$ sensor at $k_f=50$ sec during the climbing mode and the concurrent $5\%$ bias fault applied to the $N$ sensor at $k_f=250$ sec during the cruise mode. The estimated concurrent faults are depicted in \figurename{} \ref{FID_Conc1}.

The concurrent faults for the first scenario are  detected and isolated after $1.7$ and $2.4$ seconds, respectively from their occurrence instants, whereas the percentages of the WMSNEs are $0.0208\%$ and $0.106\%$, respectively. The second scenario simulates the effects of a $4\%$ bias fault that is applied to the $T_{\rm T}$ sensor at $k_f=250$ sec during the cruise mode and the concurrent $6\%$ bias fault that is applied to the $P_{\rm T}$ sensor at $k_f=450$ sec during the landing mode.
The concurrent faults for the second scenario are  detected and isolated after $0.9$ and $0.4$ seconds, respectively from their occurrence instants, whereas the percentages of the WMSNEs are $0.0125\%$ and $0.1702\%$, respectively.
\vspace{-2mm}
\subsection{\underline{Case 6: Comparison}}
In this section, the efficiency of our proposed MM-based sensor FDI scheme which utilizes the MHKF as a detection filter is compared with different filtering methods including the linear Kalman filter (LKF), extended Kalman filter (EKF), unscented Kalman filter (UKF), and the cubature Kalman filter (CKF) \cite{Arasaratnam2009}. It must be noted that in the following conducted experiments, multiple combined residual vectors and covariance matrices associated with multiple operating points are computed for both the HKF and LKF filtering methods to construct the multiple HKF (MHKF) and multiple LKF (MLKF)-based FDI schemes, respectively. Moreover, our MM-based FDI scheme is compared with the \textit{interacting multiple model} (IMM) approach \cite{Zhan1999, Ru2008}. Different experiments are conducted in this section to compare the promptness of the fault detection and isolation scheme as well as the degree of robustness towards the health parameters degradations among the above filtering methods. In all the experiments, similar process and measurement noise signals are applied to the gas turbine engine and similar noise covariance matrices are used to construct the above filters. The sampling rates for simulating the actual engine and the nonlinear filtering approaches are 0.01 sec and 0.1 sec, respectively. Also, $2n+1$ sigma points, where $n$ is the dimension of the state variable vector, are generated in the UKF method with the tuning parameters $\alpha=10^{-3}$, $\beta=2$ and $\kappa=0$. For the CKF method, $2n$ cubature points are generated through intersecting the unit sphere with the Cartesian axes and are then scaled by $\sqrt{n}$.
Due to space limitations, the details regarding the formulations of the above applied filtering methods are not included as they can be found in the literature. \\
\begin{figure*}
\centering
\includegraphics[width=0.97\textwidth]{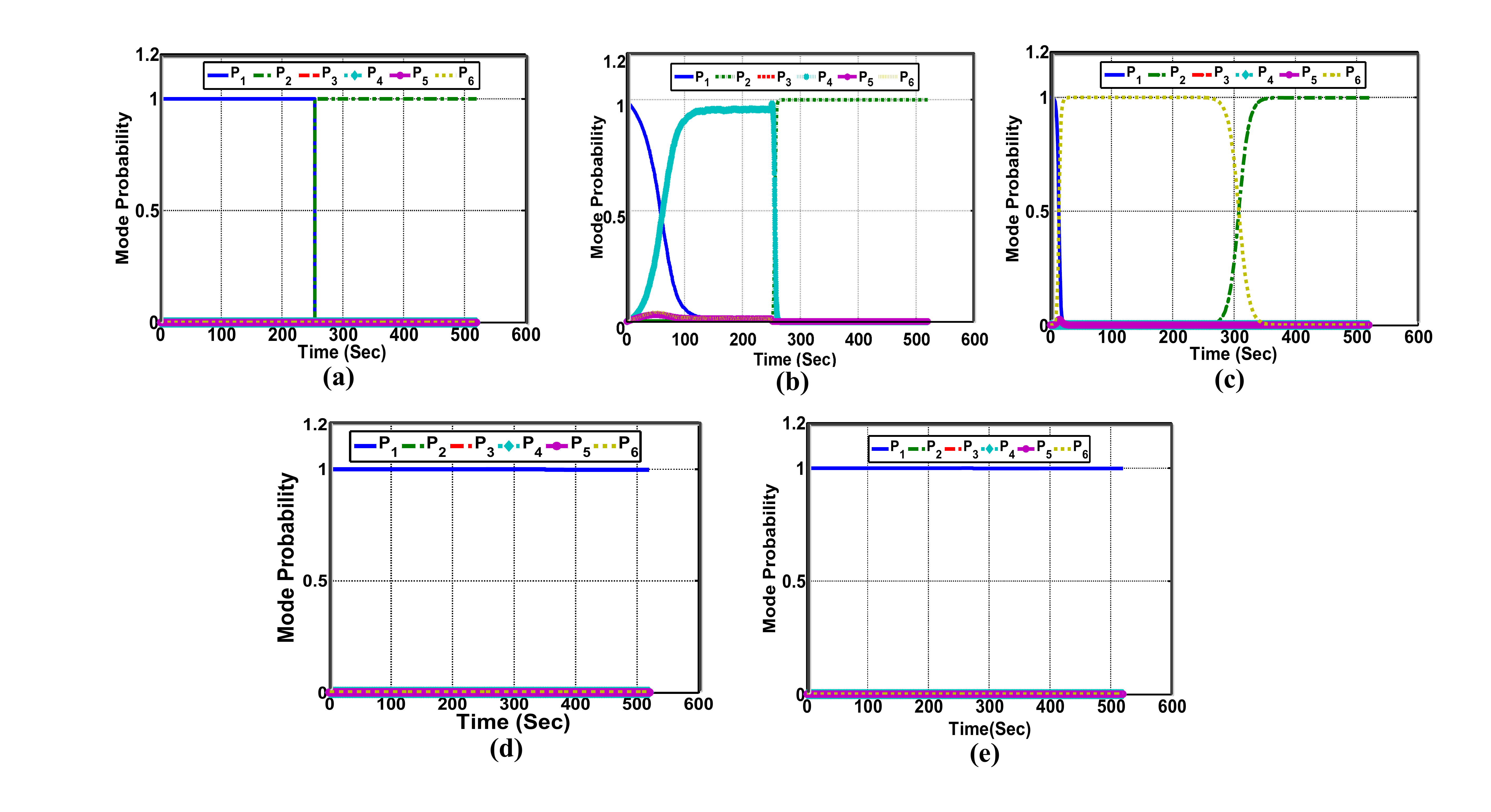}
\vspace{-4mm}
\caption{Mode probabilities for a $3\%$ bias fault that is applied at $k_f=250$ sec to the $T_{\rm C}$ sensor by applying multiple (a) HKF, (b) LKF, (c) EKF, (d) UKF and (e) CKF based FDI approaches with $2.5\%$ degradations applied to the compressor health parameters of the actual engine.}
\label{Comparison_Deg}
\vspace{-4mm}
\end{figure*}

\textbf{\underline{Experiment 1:}} The $3\%$ single bias fault is applied to different sensors at $k_f=250$ sec during the cruise mode with no health parameter degradations. All the filters are capable of detecting and isolating the injected fault correctly with different fault detection times. The sensor FDTs are shown in \tablename {} \ref{Comp} for the detection filters that are utilized in the MM-based FDI scheme. It can be concluded from \tablename {} \ref{Comp} that our proposed approach is capable of detecting and isolating the sensor faults faster than other methods, particularly the MLKF method, given that the OBEM can capture the nonlinear behaviour of the engine more effectively than others. Although, the EKF, UKF, and CKF are well-known nonlinear filtering methods, however they detect and isolate the injected sensor faults with more delay than our proposed method. According to our observations that are included in \tablename {} \ref{Error}, the HKF has smaller estimation error as compared to the EKF, UKF, and CKF, which leads to faster FDI performance. It is also expected that in case the OBEM is subjected to certain unmodeled dynamics with respect to a real engine, the FDTs that are obtained with multiple EKF, UKF, and CKF based FDI schemes are more  comparable with our proposed MHKF-based FDI scheme. \\
\begin {table}	
	\centering
	\small
	\caption{Absolute mean of residual signals associated with various types of nonlinear filters corresponding to the healthy mode in the MM-based scheme for the healthy engine sensors without health degradation.}
	\begin{tabular}{|c|c|c|c|c|} \hline
	\multirow{2}{*}{Sensor} & \multicolumn{4}{c|}{Absolute Mean of Residual Signal} \\ \cline{2-5}
 & \text{MHKF}  & \text{EKF}  & \text{UKF}  & \text{CKF}   \\ \hline
	$T_{\text C}$ & $8.8329\times 10^{-4}$ & $0.00063$ & $0.0058$ &$0.0039$ \\ \hline
	$P_{\text{C}}$ &$0.0044$ &$0.0186$ &$0.0127$ &$0.0105$ \\ \hline
	$N$ &$4.0454 \times 10^{-4}$ &$0.0032$ &$0.0011$ &$0.0009$ \\ \hline
	$T_{\text T}$ &$8.1498\times 10^{-4}$ &$0.0081$ &$0.0085$ &$0.0069$ \\ \hline
	$P_{\text T}$ & $0.0014$ & $0.0028$ & $0.0082$ & $0.0074$ \\ \hline
	\end{tabular}
\label{Error}
\end{table}

\textbf{\underline{Experiment 2:}} The robustness of our developed MHKF-based FDI approach is compared with the other filtering methods here. The $3\%$ bias fault is applied to the $T_{\rm C}$ sensor at $k_f=250$ sec during the cruise mode with $2.5\%$ degradations that are applied to the compressor health parameters. In this experiment, it is assumed that the OBEM health parameters are not updated and the $2.5\%$ degradation is only applied to the compressor health parameters of the actual engine. Therefore, the reference baselines associated with the compressor health parameters are one for both the OBEM model and the nonlinear model that is used for deriving the other nonlinear filtering methods. The mode probabilities that are generated in this scenario are depicted in \figurename {} \ref{Comparison_Deg}. It can be concluded that our method outperforms the other approaches in terms of robustness with respect to the health parameters degradations. It is also shown in \figurename {} \ref{Comparison_Deg} that the applied fault can be detected by using the EKF and MLKF methods in addition to our method, but  false alarms are generated by these methods before the occurrence of the fault.\\

\begin{figure}
\centering
\vspace{-2mm}
\includegraphics[width=\columnwidth]{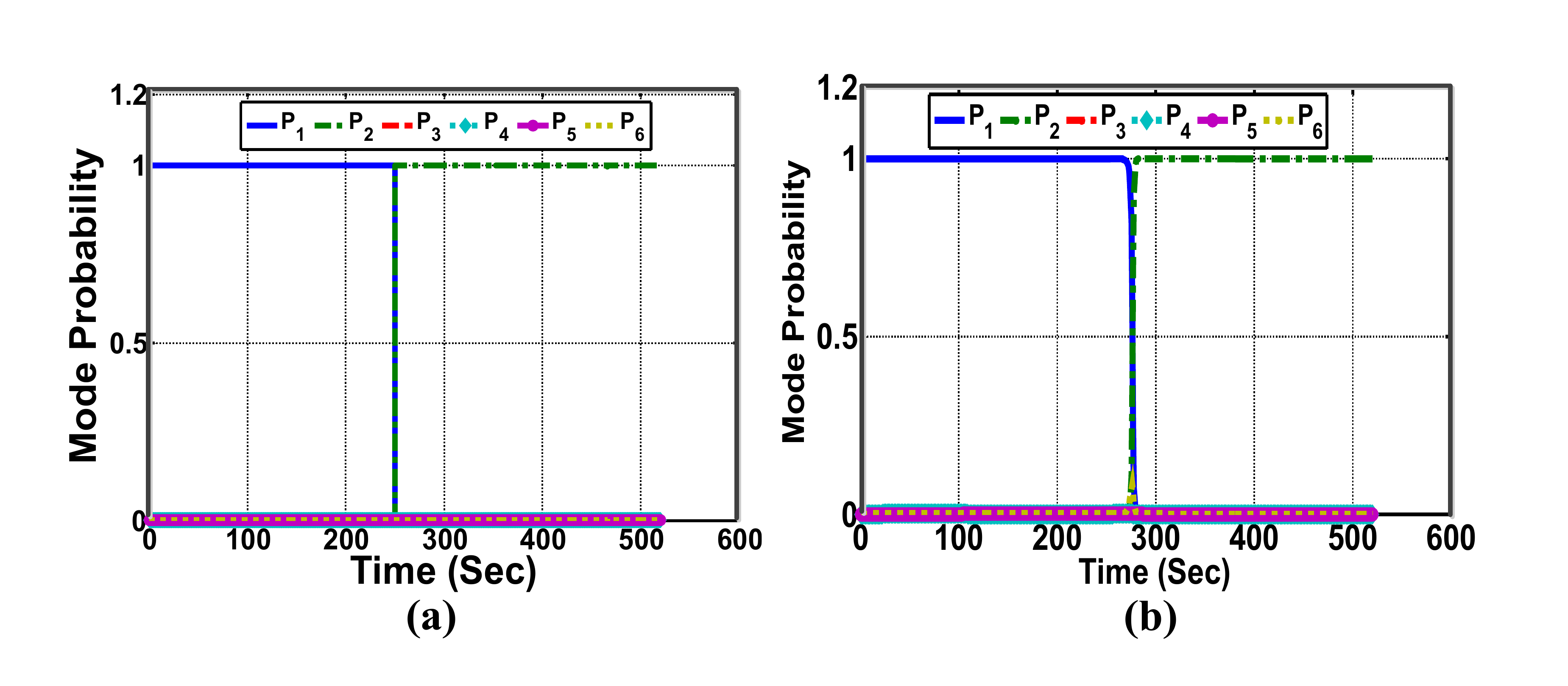}
\vspace{-7mm}
\caption{Mode probabilities for a $3\%$ bias fault applied at $k_f=250$ sec to the $T_{\rm C}$ sensor using (a) the MHKF and (b) the EKF-based FDI methods, with $\lambda=0.99$ and $\hat\lambda=0.975$.}
\label{HKF_EKF_Deg}
\vspace{-2mm}
\end{figure}

\textbf{\underline{Experiment 3:}} Our proposed MHKF method is similar to the EKF in the sense of the linearization process at certain operating points instead of the entire flight profile. Therefore, one may expect that the EKF method can yield similar results to that of the MHKF method in case the health parameters are also updated for the nonlinear engine model used for deriving the EKF. In this experiment, the same engine model where its health parameters are updated is used for the MM-based structures that utilize the MHKF and EKF schemes. The $3\%$ bias fault is applied to the $T_{C}$ sensor at $k_f=250$ sec during the cruise mode with $\lambda=0.99$ and $\hat\lambda=0.975$. The estimated health parameters are updated similarly for both the EKF and MHKF schemes. \figurename {} \ref{HKF_EKF_Deg} shows the mode probabilities that are generated in this scenario. According to \figurename {} \ref{HKF_EKF_Deg}, the EKF is also capable of correctly detecting and isolating the applied fault by using the health parameters updating mechanism. However, the fault detection time and the computational time of the EKF is far greater than that of the MHKF. The MM-based FDI method is capable of detecting and isolating the applied fault in $2.7$ sec and $20.8$ sec by using the MHKF and EKF methods, respectively. The same results are also obtained for the other fault scenarios, but are not included here due to the space limitations. \\

\textbf{\underline{Experiment 4:}} In this experiment, the performance of the FDI algorithm that utilizes our proposed MHKF as well as the other filtering methods are compared in the \ul{interacting multiple model (IMM)} structure. 
For the IMM-based FDI structure, the sensor FDTs are computed for the $3\%$ single bias fault that is applied to the sensors with no health parameter degradations as represented in the \tablename {} \ref{Comp_IMM}. According to the simulation results, the fault is detected sooner by using the IMM structure that utilizes all the detection filters except the MHKF. It appears that the interactions among the multiple models do not improve the MHKF-based FDI scheme performance. For simulating the IMM scheme which utilizes our proposed MHKF, the off-diagonal elements of the transition probability matrix should be selected less than 0.001; otherwise, the probability of the healthy mode monotonically decreases to reach almost 0.8 before the occurrence of a sensor fault which can enhance the risk of a false alarm. Therefore, it is not recommended to use the IMM scheme for the MHKF method. \\

\begin {table}
	\caption{Sensor fault detection times using multiple \underline{non-interacting} HKF, LKF, EKF, UKF and CKF based FDI approaches without considering health parameter degradation.}	
	\label{Comp}
	\centering
	\small
	\vspace{-1mm}
	\begin{tabular}{|c|c|c|c|c|c|} \hline
	\text{MM-Based} & \multicolumn{5}{c|}{Fault Detection Time (FDT) (sec)} \\ \cline{2-6}
\text{FDI Approach}& $T_{\rm C}$ & $P_{\rm C}$ & $N$ & $T_{\rm T}$ & $P_{\rm T}$ \\ \hline
	\text{HKF} &$0.2$ &$0.2$ &$0.2$ &$0.2$ &$0.3$ \\ \hline
	\text{LKF} &$8.3$ &$38.2$ &$16.9$ &$12.1$ &$12.4$ \\ \hline
	\text{EKF} &$1$ &$1$ &$0.9$ &$0.8$ &$0.9$ \\ \hline
	\text{UKF} &$0.6$ &$0.7$ &$0.6$ &$0.7$ &$2.7$ \\ \hline
	\text{CKF} &$0.6$ &$0.6$ &$0.7$ &$0.7$ &$2.3$ \\ \hline
	\end{tabular}
	\vspace{-2mm}
\end{table}

\begin {table}
	\caption{Sensor fault detection times using multiple \underline{interacting} HKF, LKF, EKF, UKF and CKF based FDI approaches without considering any health parameter degradation.}	
	\label{Comp_IMM}
	\centering
	\small
	\vspace{-1mm}
	\begin{tabular}{|c|c|c|c|c|c|} \hline
	\text{MM-Based} & \multicolumn{5}{c|}{Fault Detection Time (FDT) (sec)} \\ \cline{2-6}
\text{FDI Approach} & $T_{\rm C}$ & $P_{\rm C}$ & $N$ & $T_{\rm T}$ & $P_{\rm T}$ \\ \hline
	\text{HKF} &$0.3$ &$0.4$ &$0.3$ &$0.3$ &$0.5$ \\ \hline
	\text{LKF} &$6.3$ &$35.8$ &$15.7$ &$11.6$ &$11.8$ \\ \hline
	\text{EKF} &$0.9$ &$0.9$ &$0.7$ &$0.8$ &$0.8$ \\ \hline
	\text{UKF} &$0.5$ &$0.5$ &$0.5$ &$0.6$ &$2.3$ \\ \hline
	\text{CKF} &$0.5$ &$0.5$ &$0.5$ &$0.5$ &$1.9$ \\ \hline
	\end{tabular}
	\vspace{-2mm}
\end{table}
\textbf{\underline{Experiment 5:}} In this experiment, the computational time of all the above filtering methods are obtained and compared with our proposed MHKF approach in \tablename{} \ref{Time}. This comparison confirms the advantages of implementing the MHKF approach in real-time applications. The computational time is measured for only the filters corresponding to the healthy mode by the use of \emph{``tic-toc''} command in Matlab. It must be noted that the computational time of the MHKF includes the time for not only simulating the healthy mode but also for running the OBEM model. Therefore, based on the information provided in \tablename {} \ref{Time}, the MHKF scheme is faster and more suitable than the other filtering methods to be utilized during the entire flight profile. Computing the Jacobian functions in each time step for the EKF method and also the cubature and sigma points with their corresponding mean and covariance weight matrices for the CKF and UKF methods, respectively are time consuming operations that lead to the higher computational time compared to our proposed MHKF scheme. Moreover, the MLKF approach computes the Kalman gain matrices on-line, whereas the MHKF utilizes the previously stored gain matrices. However, if the same steady-state Kalman gain matrices are computed off-line and stored for use in the MLKF approach, the average computational time associated with the healthy mode filters in the MLKF scheme is about 80 seconds, which is less than that of the MHKF method. Under this situation, the extra computational time required by the MHKF is associated with the OBEM operations. Therefore, the only disadvantage of the MHKF method is the amount of memory that is needed for storing the look-up tables that include the matrices $A^i$, $C^i$ and $K_{\rm ss}^i$ at different operating points. 

Finally, in the following experiment results, it is shown that the number of required operating points, and therefore the amount of required memory for the MHKF, is less than that of the MLKF-based FDI method. \\
\begin {table}
	\caption{Computational time as measured in seconds for different filtering methods.}	
	\label{Time}
	\centering
	\small
	\vspace{-1mm}
	\begin{tabular}{|c|c|c|c|c|} \hline
	\text{MHKF} & \text{MLKF} & \text{EKF} & \text{UKF} & \text{CKF} \\ \hline
	$131.21$ & $526.81$ & $1184.2$ & $2856.4$ & $2673.44$ \\ \hline
	\end{tabular}
	\vspace{-3mm}
\end{table}

\begin{figure*}
   \centering
   \includegraphics[width=0.97\textwidth]{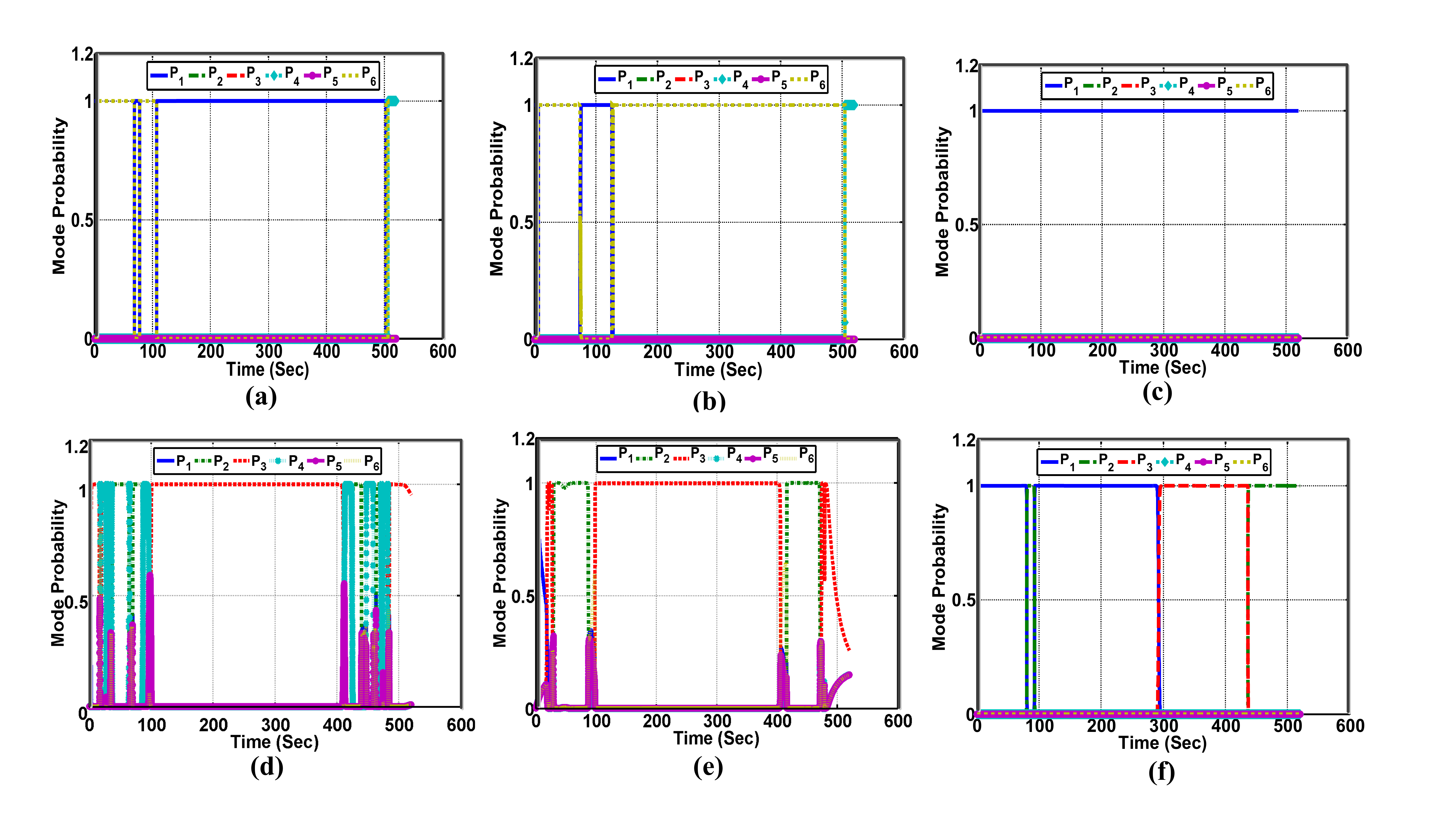}
   \vspace{-5mm}
   \caption{Mode probabilities that are generated for the MM-based FDI method by utilizing (a-c) the MHKF and (d-f) the MLKF with one, three and five operating points.}
   \label{LKF_HKF_PWL}
\end{figure*}

\textbf{\underline{Experiment 6:}} In this experiment, the number of required operating points is investigated by measuring the mean of the combined residual signals in presence of the health parameter degradations for the MLKF and MHKF methods. To cover the entire flight profile, five operating points are generated for the MHKF method as described in \tablename {} \ref{tbl:PWL}. Note that the MLKF method still needs more operating points since it is not capable of tracking rapid variations of the thrust during the climbing and landing modes, and therefore it generates false alarms due to the higher estimation error as compared to the MHKF scheme. \tablename {} \ref{Error_LKF_HKF} compares the mean of the MHKF and MLKF combined residual signals that are associated with the healthy mode as a function of the number of operating points. \figurename {} \ref{LKF_HKF_PWL} displays the mode probabilities that are generated for the MHKF and MLKF methods for different number of operating points. In this experiment, the engine has no sensor fault, but a $3\%$ degradation is applied to the actual engine compressor health parameters. As stated in Section \ref{False Alarm}, the MHKF is robust with respect to the applied health degradations even without updating the OBEM model health parameters, although the MHKF needs to be updated for larger degradation levels. Based on the results obtained, it can be concluded that the mean of the combined residual signal associated with the healthy mode can be reduced by increasing the number of the operating points. It is expected that the MLKF estimation error decreases further by increasing the number of the operating points. However, we are not capable of perfectly avoiding generation of a false alarm by using the MLKF method corresponding to our particular flight profile by increasing the number of the operating points, since the MLKF cannot track high variations of the thrust in our application. Therefore, it may be feasible to apply the MLKF with the combined PWL models if the engine thrust increases with the slower speed than ours since the PWL model weights calculation is a time-consuming process which may diminish the ability of the MLKF to track the rapid variations of the thrust.

\begin {table}	
	\caption{Mean of the combined residual signals $(\overline{\gamma}_c^1)$ associated with the healthy mode for both the MHKF and MLKF methods in terms of the number of the operating points, whereas $3\%$ degradation is applied to the compressor health parameters.}	
    \centering
	\label{Error_LKF_HKF}
	\scalebox{0.92}{
	\small{
	\vspace{-1mm}
	\hspace{-3mm}
	\begin{tabular}{|c|c|c|c|c|c|c|} \hline
		\multirow{2}{*}{\text{Sensor}} & \multicolumn{2}{c|} {\text{1 operating point}}& \multicolumn{2}{c|} {\text{3 operating points}} & \multicolumn{2}{c|} {\text{5 operating points}}\\ \cline{2-7}
		 & \text{HKF} &	\text{LKF}	& \text{MHKF} & \text{MLKF} & \text{MHKF} & \text{MLKF}  \\ \hline
	  $T_{\rm C}$ & $0.109$ & $0.2305$ & $0.077$ & $0.1091$ & $0.0119$ & $0.0857$ \\ \hline
	  $P_{\rm C}$ & $0.2286$ & $0.943$ & $0.0771$ & $0.1523$ & $0.0216$ & $0.1315$  \\ \hline	
 	  $N$ & $0.0255$ & $0.0123$ & $0.0167$ & $0.0053$ & $0.0003$ & $0.0014$  \\ \hline	
 	  $T_{\rm T}$ & $0.0813$ & $0.3339$ & $0.0745$ & $0.2372$ & $0.0197$ & $0.1008$  \\ \hline	
 	  $P_{\rm T}$ & $0.0152$ & $0.0477$ & $0.0116$ & $0.036$ & $0.0006$ & $0.0049$  \\ \hline	
	\end{tabular}}}
\end{table}
Therefore, it can be concluded that the MM-based FDI scheme which utilizes our proposed MHKF approach is capable of promptly detecting and isolating various sensor bias faults during the \underline{entire flight profile} by updating the OBEM health parameters reference baselines and even without updating the OBEM model for certain levels of health parameter degradations. Our scheme also estimates the engine state and output variables with more accuracy and smaller number of operating points as compared to that by the MLKF-based scheme.

\vspace{-1.5mm}
\section{CONCLUSIONS}
\vspace{-1mm}
In this paper, a novel approach for both single and concurrent sensor fault detection, isolation and identification/estimation (FDII) for gas turbine engines is proposed. Our methodology is based on the developed \emph {Hybrid Kalman Filter (HKF)} as the detection filter of a \emph{hierarchical multiple-model} based structure. Despite the use of linear Kalman filters, the HKF is capable of capturing the nonlinearities of the system by integrating a nonlinear on-board engine model (OBEM) with  piecewise linear (PWL) models to cover the entire operating range of the engine. Compared to the multiple linear Kalman filter (MLKF), our proposed approach requires fewer number of operating points although each corresponds to a larger operating range.

Another important contribution of this work is in the inclusion of the effects of health parameter degradations in our proposed sensor FDII scheme through updating the OBEM health parameters reference baselines that enables one to prevent false alarms and incorrect fault detections.
Therefore, unlike most of the previous work in the literature that have not considered the influence of health parameters variations on the performance of the developed FDII approaches, our proposed sensor FDII scheme is sufficiently more effective for use in the entire flight profile. It is furthermore suitable during the engine life cycle by updating the OBEM health parameters reference baselines, and by replacing the steady-state values with the OBEM states and outputs and integration of the designed PWL models. Moreover, our proposed MM-based sensor FDI scheme is integrated with the modified GLR method to estimate a sensor fault severity.
Finally, it is demonstrated through extensive simulation studies that the accuracy and robustness of our proposed MHKF-based FDI approach against the engine health parameter degradations are significantly superior to that of those generated by other investigated methods such as the MLKF, EKF, UKF, and CKF.

\appendices
\vspace{-2mm}
\section{}\label{Appendix A}
In Section \ref{HKF}, for sake of notational simplicity the same notations were used for both versions of the OBEM. However, here we use $X_{\rm NOBEM}$ and $Y_{\rm NOBEM}$ to designate the nominal OBEM $(\lambda=1)$ that is equivalent to \eqref{2}, and $X_{\rm OBEM}$ and $Y_{\rm OBEM}$ to designate the OBEM with the updated health parameters $(\hat\lambda\neq 1)$ that is equivalent to \eqref{5}.

Let the nominal continuous-time OBEM model (NOBEM) be represented by:
\begin{align}
&\dot{X}_{\rm{NOBEM}}(t)=f_c(X_{\rm{NOBEM}}(t),H_{\rm NOBEM}(t),U(t)), \notag\\
&Y_{\rm{NOBEM}}(t)=g_c(X_{\rm{NOBEM}}(t),H_{\rm NOBEM}(t)).\notag
\end{align}

To construct the HKF, the nominal continuous-time OBEM is initially linearized at a certain operating point $(X_{\rm ss},U_{\rm ss},Y_{\rm ss})$ without considering degradations as follows:
\begin{align}
&\dot{X}_{\rm{NOBEM}}(t)=f_c(X_{\rm{ss}},U_{\rm{ss}})+\frac{\partial f_c}{\partial X_{\rm{NOBEM}}}|_{X_{\rm{ss}}}(X_{\rm{NOBEM}}(t)-X_{\rm{ss}})+\frac{\partial f_c}{\partial U}|_{U_{\rm{ss}}}(U(t)-U_{\rm{ss}})+\Delta f_c, \notag\\
&Y_{\rm{NOBEM}}(t)=g_c(X_{\rm{ss}},U_{\rm{ss}})+\frac{\partial g_c}{\partial X_{\rm{NOBEM}}}|_{X_{\rm{ss}}}(X_{\rm{NOBEM}}(t)-X_{\rm{ss}})+\Delta g_c,\notag
\end{align}

\noindent where $f_c(X_{\rm{ss}},U_{\rm{ss}})\equiv 0$ and $g_c(X_{\rm{ss}},U_{\rm{ss}})\equiv Y_{\rm ss}$. The health parameters are the state-dependent functions whose effects are incorporated in the above partial derivatives. The linear continuous-time model is now discretized by using a sufficiently small sampling period to yield the corresponding discrete-time linear model:
\begin{align}
\vspace{-2mm}
&X_{\rm NOBEM}(k+1)-X_{\rm ss} =A|_{X_{\rm ss}}(X_{\rm NOBEM}(k)-X_{\rm ss})+B|_{U_{\rm ss}}(U(k)-U_{\rm ss})+\Delta F, \notag\\
&Y_{\rm NOBEM}(k)=Y_{\rm ss}+C|_{X_{\rm ss}}(X_{\rm NOBEM}(k)-X_{\rm ss})+ \Delta G,\notag
\vspace{-4mm}
\end{align}

\noindent where $A|_{X_{\rm ss}}={\rm exp}({{A_c}|_{X_{\rm ss}}T_s})$, $B|_{U_{\rm ss}}=(\int_0^{T_s}{\rm exp}({{A_c}|_{X_{\rm ss}}\tau}){\rm d}\tau){B_c}|_{U_{\rm ss}}$, and $C|_{X_{\rm ss}}={C_c}|_{X_{\rm ss}}$ are obtained in terms of the state-space matrices that are associated with the continuous-time linear model ${A_c}|_{X_{\rm ss}}=\frac{\partial f_c}{\partial X_{\rm{NOBEM}}}|_{X_{\rm{ss}}}$, ${B_c}|_{U_{\rm ss}}=\frac{\partial f_c}{\partial U}|_{U_{\rm{ss}}}$, and ${C_c}|_{X_{\rm ss}}=\frac{\partial g_c}{\partial X_{\rm{NOBEM}}}|_{X_{\rm ss}}$, and the sampling period is denoted by $T_s$. Also, $\Delta f_c$, $\Delta g_c$, $\Delta F$ and $\Delta G$ represent the higher order terms in the linearization process. To construct the HKF, it is necessary to update the OBEM health parameters. Therefore, the nominal model is only used once to derive the state-space matrices and the steady-state values.

\textbf{Assumption 1.} The linearization of the nominal OBEM with $\lambda=1$ and the updated OBEM with $\hat\lambda\neq1$ generate approximately the same matrices and steady-state values.

Therefore, given that Assumption 1 holds, the linearized version of the updated OBEM can be derived as follows:
\begin{align}
\vspace{-2mm}
&X_{\rm OBEM}(k+1)- X_{\rm ss} \approx A|_{X_{\rm ss}}(X_{\rm OBEM}(k)-X_{\rm ss})+B|_{U_{\rm ss}}(U(k)-U_{\rm ss})+\Delta F, \notag\\
&Y_{\rm OBEM}(k)\approx Y_{\rm ss}+C|_{X_{\rm ss}}(X_{\rm OBEM}(k)-X_{\rm ss})+ \Delta G.
\label{AP4}
\vspace{-4mm}
\end{align}

Moreover, the relationship between the instantaneous values of $X_{\rm{OBEM}}(k)$ and $Y_{\rm{OBEM}}(k)$ and the steady-state values associated with each linearized model can be given by:
\begin{align}
& X_{\rm{OBEM}}(k)=X_{\rm{ss}}+X_l(k),\notag\\
& Y_{\rm{OBEM}}(k)=Y_{\rm{ss}}+Y_l(k),
\label{AP5}
\vspace{-1mm}
\end{align}

\noindent where $X_l(k)$ and $Y_l(k)$ represent the perturbations from the steady-state values as well as variations of the OBEM state variables and outputs due to updating the health parameters reference baselines. The linear Kalman filter can be designed as follows:
\begin{align}
&\hat{X}(k+1)-X_{\rm ss}=A|_{X_{\rm ss}}(\hat{X}(k)-X_{\rm{ss}})+B|_{U_{\rm ss}}(U(k)-U_{\rm{ss}})+ K_{\rm ss}(Y(k)-\hat{Y}(k)), \notag\\
&\hat{Y}(k)=Y_{\rm{ss}}+C|_{X_{\rm ss}}(\hat{X}(k)-X_{\rm{ss}}),\notag
\end{align}

\noindent which can be rewritten by using \eqref{AP5} as follows:
\begin{align}
\hat{X}(k+1)-X_{\rm OBEM}(k)+X_l(k)=A|_{X_{\rm ss}}(\hat{X}(k)-X_{\rm{OBEM}}(k))&+K_{\rm ss}(Y(k)-\hat{Y}(k))+A|_{X_{\rm ss}}X_l(k)\notag\\
&+B|_{U_{\rm ss}}(U(k)-U_{\rm{ss}}), \notag\\
\hat{Y}(k)=Y_{\rm{OBEM}}(k)+C|_{X_{\rm ss}}(\hat{X}(k)-X_{\rm{OBEM}}(k))+C|_{X_{\rm ss}}X_l(k)&-Y_l(k).
\label{AP7}
\end{align}

Using \eqref{AP4}, the last two terms in the R.H.S of \eqref{AP7} can be rewritten as follows:
\begin{align}
&A|_{X_{\rm ss}}(X_{\rm{OBEM}}(k)-X_{\rm{ss}})+B|_{U_{\rm ss}}(U(k)-U_{\rm{ss}})\approx X_{\rm{OBEM}}(k+1)-X_{\rm ss}-\Delta F, \notag\\
&C|_{X_{\rm ss}}(X_{\rm{OBEM}}(k)-X_{\rm{ss}})-Y_l(k)\approx Y_{\rm{OBEM}}(k)-Y_{\rm{ss}}-Y_l(k)-\Delta G=-\Delta G.\notag
\end{align}

Therefore, the discrete-time form of the HKF can be obtained as follows:
\begin{align}
&\hat{X}(k+1)-X_{\rm{OBEM}}(k+1)\approx A|_{X_{\rm ss}}(\hat{X}(k)-X_{\rm{OBEM}}(k))+K_{\rm ss}(Y(k)-\hat{Y}(k))-\Delta F, \notag\\
&\hat{Y}(k)\approx Y_{\rm{OBEM}}(k)+C|_{X_{\rm ss}}(\hat{X}(k)-X_{\rm{OBEM}}(k))-\Delta G,\notag
\end{align}

\noindent which will lead to \eqref{6} for the particular operating point if the higher order terms $\Delta F$ and $\Delta G$ as well as the approximation error due to Assumption 1 are neglected.

\bibliographystyle{IEEEtr}
\bibliography{IEEEabrv,References_new}
\end{document}